\documentclass[referee]{raa}   
\usepackage{graphicx,times}
\usepackage{natbib}
\usepackage{amssymb,amsmath}
\bibpunct{(}{)}{;}{a}{}{,}

\usepackage[pagebackref=true]{hyperref}

\def\be{\begin{eqnarray}}\def\ee{\end{eqnarray}}

\def\lsim{\mathrel{\rlap{\lower3pt\hbox{\hskip1pt$\sim$}}
		\raise1pt\hbox{$<$}}} 
\def\gsim{\mathrel{\rlap{\lower3pt\hbox{\hskip1pt$\sim$}}
		\raise1pt\hbox{$>$}}} 

\newcommand\sect[1]{\section{#1}}

\begin{document}
	
\title{ Probing Hadron-quark Transition Through Binary Neutron Star Merger}
\volnopage{ {\bf 20XX} Vol.\ {\bf X} No. {\bf XX}, 000--000}
\setcounter{page}{1}

\author{
		Ling-Jun Guo\inst{1,2}, Wen-Cong Yang\inst{1,2}, Yong-Liang Ma\inst{2,3,4,5},Yue-Liang Wu\inst{4,5,6}
	}
\institute{ Center for Theoretical Physics and College of Physics, Jilin University, Changchun 130012, 
China; \\
        \and
		School of Fundamental Physics and Mathematical Sciences,
		Hangzhou Institute for Advanced Study, UCAS, Hangzhou 310024, China\\
	\and
School of Frontier Sciences, Nanjing University, Suzhou 215163, China; {\it ylma@nju.edu.cn}\\
\and
TaiJi Laboratory for Gravitational Wave Universe (Beijing/Hangzhou), University of Chinese Academy of Sciences, Beijing 100049, China; {\it ylwu@ucas.ac.cn}\\
\and
International Center for Theoretical Physics Asia-Pacific (ICTP-AP) , UCAS, Beijing 100190, China\\
\and
Institute of Theoretical Physics, Chinese Academy of Sciences, Beijing 100190, China
\vs \no
   {\small Received 20XX Month Day; accepted 20XX Month Day}
}




\date{\today}

\abstract{
The cores of massive neutron stars offer a unique environment for the nuclear matter at intermediate density in the universe. The global characteristics of a neutron star, 
as well as the gravitational waves emitted from the mergers of two neutron stars, offer valuable insights into dense nuclear matter. In this paper, 
we comprehensively investigate the effect of the potential hadron-quark transition on the properties of neutron stars and the signals of the gravitational waves stemming 
from the merger of binary neutron stars, including waveforms, frequency evolutions as well as the spectrum curves, utilizing the equations of state constructed from the Maxwell ansatz, 
Gibbs ansatz and, the crossover scenario. We explicitly construct the equations of state in such a way that they converge at low and high densities therefore the differences 
are only from the scenarios of the transitions and the locations---or the parameters in the equation of state. Using such constructed equations of state, 
we simulate the signals of the gravitational wave (GW) and analyze their differences due to locations of the transition, the scenarios of the transition, and the masses of the component stars. We find that (1) in both the Maxwell ansatz and Gibbs ansatz, GW signals are sensitive to the location and the latent heat of the phase transition, 
(2) in the post-merger phase, the frequency of GW increases with the evolution in Maxwell type transition but is stable in the other two types of transitions and, (3) the amount of radiated energy is the biggest in Gibbs construction (GC) type transition and the smallest in the crossover construction (CC) type transition. By combining our findings with the expected detection of gravitational waves around $(2$-$4)$ kHz from binary neutron star mergers and their associated electromagnetic signals, 
we expect to uncover some key characteristics of dense nuclear matter.
}


\authorrunning{L.-J. Guo et al. }            
\titlerunning{Probing hadron-quark transition through binary neutron star merger}  
\maketitle

\sect{Introduction}

Despite decades of study, there are still numerous fundamental questions regarding dense nuclear matter at low temperature,  awaiting clarification~\cite{Brown:2001nh,Subedi:2008zz,Fukushima:2013rx,Holt:2014hma,Drews:2016wpi,Baym:2017whm,Ma:2019ery,Li:2019xxz,Annala:2019puf,Lovato:2022vgq}. Its properties at densities $n \gsim 2n_{0}$, with $n_0 \simeq 0.16$~fm$^{-3}$ being the saturation density, cannot be accessed by terrestrial experiments or Lattice QCD simulation. Due to the strong gravitational force, neutron star (NS) provides a unique environment for understanding the cold nuclear matter at densities ranging from $\approx 2n_0$ to approximately $10 n_0$ in the universe.

The observations of the massive pulsars~\cite{Demorest:2010bx,Antoniadis:2013pzd,Fonseca:2016tux,NANOGrav:2017wvv,NANOGrav:2019jur,Riley:2019yda,Fonseca:2021wxt,Riley:2021pdl} have effectively ruled out the soft equations of state (EoSs) of nuclear matter, which are unable to produce an NS with a mass $\simeq 2.0M_\odot$ with $M_\odot$ being the solar mass. The joint constraints from the global properties of massive NSs and data heavy-ion collisions further narrowed the bound of EoS~\cite{Klahn:2006ir,Tsang:2018kqj}. Recently, the detection of the gravitational waves (GWs) emitted during the merger of binary neutron stars (BNSs) at LIGO-Virgo collaboration in GW170817 has provided a novel means for investigating dense nuclear matter~\cite{LIGOScientific:2017vwq,Orsaria:2019ftf}, supplementing the use of the electromagnetic probes. The GW170817 has provided additional constraints on the EoS of nuclear matter through the extracted tidal deformability---a quantity sensitive to the EoS that describes how much a star is deformed in the presence of a tidal field~\cite{Annala:2017llu,Tews:2018iwm}. Furthermore, the evolution of the GWs emitted from merger processes carries the information of dense nuclear matter, as the merger process inevitably involves the exchange of matter.

In the low density region up to $\sim2.0n_0$, the EoS of nuclear matter can be well described by the pure hadronic models and well constrained by the terrestrial experiments. At super-high density $\gsim 100 n_0$, the EoS can be reliably calculated by using the first principle of strong interaction---quantum chromodynamics (QCD)~\cite{Freedman:1976ub}. However, within the intermediate density region relevant to the cores of compact stars up to $\sim 10n_0$, the EoS presents a complex and challenging scenario~\cite{Chen:2015gba,Gil:2018yah,Li:2019xxz}. Typically, the compact star matter is described in terms of pure hadrons, quark-hadron transitions, quarkyonic matter, hyperons or topology objects~\cite{Holt:2014hma,Fukushima:2013rx,Lim:2015lia,Baym:2017whm,Ma:2019ery,Oter:2019rqp,McLerran:2018hbz}, with no definitive resolution among these models yet emerging. Although the inclusion of additional degrees of freedom, such as quarks and hyperons, in addition to nucleons, softens the EoS, their presence cannot be definitely excluded and is still consistent with the observation of the massive NSs with masses $\simeq 2.0 M_{\odot}$~\cite{Kanakis-Pegios:2020jnf,Han:2018mtj,Glendenning:1991es,Lalazissis:1996rd,Han:2020adu,Lopes:2021zfe}. In the literature, the BNS merger processes have been extensively explored using the EoSs derived from the models with only hadrons~~\cite{Hotokezaka:2011dh,Hotokezaka:2013iia,DePietri:2015lya,Maione:2016zqz}, topology change effect~\cite{Yang:2020ucv}, and hyperons~\cite{Sekiguchi:2011mc,Radice:2016rys}. These studies clearly show that the GWs emitted from the post-merger phase of the BNS mergers are significantly influenced by the constituents of the cores of NSs. The present work aims to investigate the potential quark-hadron transition in the NS through the BNS merger.

The dynamical mechanism of hadron-quark transition is still an open question. No definitive evidence currently exists to conclusively determine whether the transition is of the first order or crossover along the density axis. 
When the hadron-quark transition is a first order phase transition, the energy density of the matter exhibits a distinct gap. This gap can either be independent of density or show a soft dependence, creating a soft hadron-quark mixing phase. EoS with the former characteristic is termed Maxwell-type~\cite{Margaritis:2019hfq,Kalogera:1996ci,Zdunik:2012dj,Alford:2013aca,Koranda:1996jm,Alsing:2017bbc}, while the latter is referred to as Gibbs-type~\cite{Endo:2005va,Bhattacharyya:2009fg,Fischer:2017lag}. Moreover, Lattice QCD simulations have indicated that the transition from hadrons to quark-gluon plasma at finite temperature and zero chemical potential, albeit with nonzero quark masses, is a continuous crossover~\cite{Borsanyi:2010cj,Albright:2014gva,Albright:2015uua}. Although the cold dense matter cannot be directly accessed by Lattice QCD simulation due to the sign problem, it is reasonable to contemplate the possibility of a crossover scenario for the transition at finite density within the context of neutron stars~\cite{Baym:2019iky,Kapusta:2021ney}. For the exploration of various hadron-quark transitions in dense nuclear matter, ranging from the first-order phase transition to the continuous crossover, we refer to, e.g., Refs.~\cite{Fukushima:2010bq,McLerran:2018hbz,Baym:2019iky,Fukushima:2020cmk,Kojo:2021wax}. The details of these three types of quark-hadron transitions will be discussed later.

In the literature, the effects of hadron-quark transition on the GWs from the BNS mergers have been studied through numerical relativity simulation and Bayesian analysis~\cite{Sarin:2020pwr,Mondal:2023gbf}. In Refs.~\cite{Most:2018eaw,Most:2019onn,Chatziioannou:2019yko}, a mean field EoS with the strong first-order hadron-quark phase transition, characterized by the Polyakov-loop, was utilized to study the GW signals. The authors show that the appearance of quark alters the post-merger signals considerably compared to the expected signals from the inspiral phase dominated by hadrons, potentially leading to a rapid black hole (BH) formation. It is also found that the observations of g-modes with frequencies between 1 and 1.5 kHz could be interpreted as evidence of a sharp hadron-quark phase transition in the cores of neutron stars~\cite{Orsaria:2019ftf}. Bauswein et al. studied the GW signals used the EoS featuring a mixing phase of hadrons and quarks---the Gibbs-type EoS~\cite{Bauswein:2018bma,Bauswein:2020ggy}. They found that a phase transition takes place during merger resulted in a distinctive increase in the dominant post-merger GW frequency relative to the tidal deformability characterizing the inspiral phase. By combining a low-density pure hadronic EoS with a quark matter EoS modeled using an extended bag model, Prakash et al. developed a finite temperature composition-dependent EoS with a first-order phase transition and simulated the BNS merger processes~\cite{Prakash:2021wpz}. They found that the softening of the EoS due to the phase transition led to the formation of more compact remnants and a shorter duration before collapsing to BHs. In addition, they also found that the phase transition influenced the post-merger GW signal in terms of duration, amplitude and peak frequency. In Refs.~\cite{Prakash:2021wpz,Fujimoto:2022xhv}, the authors studied the GW signals from the BNS mergers using an EoS with hadron-quark smooth crossover transition at intermediate density,  constrained by perturbative QCD (pQCD) at super-high density, and compared the results with those obtained using an EoS with a first-order phase transition. Their findings suggested that early BH collapse in the post-merger phase signals a softening of the EoS associated with the quark matter onset in the	crossover scenario. 

In this paper, we make a comprehensive study of the potential impacts of the potential hadron-quark transition on the GW signals without considering the degrees of freedom including strange quark. Instead of considering a concrete model, We will utilize the EoSs generally parameterized in terms of Maxwell construction (MC)~\cite{Zdunik:2012dj,Alford:2013aca}, Gibbs construction (GC)~\cite{Endo:2005va,Bhattacharyya:2009fg} and crossover construction (CC)~\cite{Masuda:2012kf,Kojo:2014rca} as all the effective models of EoSs can be categorized one of them. In low density region, without loss of generality, we use a typical EoS derived by a pure hadronic model for the purpose to investigate the effect of hadron-quark transition. In the super-high density region all the EoSs are constrained by pQCD. 

The main progresses distinct from the existing literature can be summarized as follows:
\begin{enumerate}
	\item When there is a hadron-quark phase transition, we explicitly compared the GW signals in the time domain and frequency domain by varying the parameters---the location of the phase transition, the energy gap relating to the energy difference between the ending and beginning of the co-existing phase---in each construction, the MC and GC, and explicitly showed that distinctions arise from the phase transition by illustrating the evolution of the EoSs covered in the BNS merger. The variation of these parameters can be attributed to the thermal effect in compact star matter~\cite{Blacker:2023afl}. If a certain type of phase transition is favored, our result would be helpful for estimating the parameters in the construction, for example, the location of the phase transition.
	
	\item The GW signals for the CC are also calculated by choosing three sets of parameters for the purpose to investigate the impact of the transition region. Instead of choosing the parameters randomly, we choose their values in each set in such a way that the EoSs from all the three sets of parameters are coincide at very low and high densities. Therefore, the effect of the central value of the crossover density region, the intensity of the transition density, the location of the transition and the energy gap can be explicitly illustrated.
	
	\item  We also compare the GW signals stemming from the BNS system with different parent masses concerning that the size of the quark core in an NS depends on the NS mass for a specific EoS. It is found that the amplitude of the GW in the inspiral phase is very sensitive to the component mass of the BNS system. 
	
	\item By suitably choosing the parameters in the MC, GC and CC type transitions such that the EoSs converge to coincidence at low and high densities but different in the intermediate densities, we compared GWs signals from these three types EoSs for the purpose of investigating the effect of the type of the transition on the GW signals. We find that the peak of the frequency spectrum can be used to distinguish the type of transition. This is one of our main qualitative conclusions in this work.
	
\end{enumerate}

To our knowledge, these factors have not been systematically simulated and compared in a single work. Although some qualitative conclusions can be naively expected, we hope our detailed analysis and comparison in this work can provide a comprehensive understanding of the phenomenon of the hadron-quark transition as well as its GW diagnosis.

The rest of this paper is arranged as follows. In Sec.~\ref{section:eos} we discuss the EoSs constructed using MC, GC, and CC. We present the details of the techniques used in this work in Sec.~\ref{sec:metho}. In Sec.~\ref{sec:GWTr} we illustrate the GWs stemming from the BNS merger by varying the parameters in each model for the purpose of exploring the parameter dependence of the GW signals. The effect of the size of the quark core in the parent stars on the GW signals was probed in Sec.~\ref{sec:CM} by varying the component mass in the binary system. We finally in sec.~\ref{sec:Type} investigate the impact the type of the hadron-quark transition on the GW signals by unifying the EoSs at low and high densities. Our conclusion and discussion are given in the last section.

\sect{Equation of state with hadron-quark transition}
\label{section:eos}
By assuming that the hadron-quark transition is of the first order, one can phenomenologically construct the hadron-quark mixed phase using a polytropic EoS of the form~\cite{1993AA268360J}
\be
p & = & \kappa n^\gamma,
\label{eq:polyeos}
\ee
where $p$ is pressure, $\gamma$ is the polytropic index, and $\kappa$ is a parameter. By applying the first law of thermodynamics, which should be fulfilled in the compact star matter
\be
p & = & n^2\frac{\partial(\varepsilon/n)}{\partial n} = n\frac{\partial\varepsilon}{\partial n} - \varepsilon,
\label{eq:thermal}
\ee
the energy density $\varepsilon$ can be obtained as a function of density based on the polytropic form~\eqref{eq:polyeos}.

When the polytropic index is set to $\gamma=0$, the system characterizes a sharp hadron-quark phase transition where pressure is independent of the matter density. This transition, referred to as the MC, is encapsulated through the following EoS~\cite{Alford:2013aca,Alford:2015dpa}
\be
\varepsilon_{\rm MC}(p) & = & \left \{
	\begin{array}{ll}			
		\varepsilon_{h}(p),                    & p\leq p_{\rm tr}\\			
		\varepsilon_h(p_{\rm tr}) + \Delta\varepsilon + (p-p_{\rm tr})c^{-2}_{\rm s},     & p\geq p_{\rm tr} \\		
	\end{array}		
	\right.	.
\label{eq:EoSMC}
\ee
Here $\varepsilon_{\rm MC}$ stands for the energy density of the MC, $\varepsilon_h$ represents the energy density derived from the pure hadronic models that will be taken, e.g., from the APR EoS~\cite{Akmal:1998cf} in this work, $p_{\rm tr}$ denotes the pressure at which the phase transition occurs, and $\Delta\varepsilon$ signifies the energy gap between the two phases. The sound speed, $c_{\rm s}$, adheres to the causality constraint $c_{\rm s}\leq 1$ (in natural unit) and is treated as a constant in the constant speed of sound (CSS) approximation. Therefore, in MC, there are three free parameters, $p_{\rm tr}, \Delta\varepsilon$ and $c_s$. In previous studies~\cite{Benic:2014jia,Alford:2017qgh,Alvarez-Castillo:2017qki,Han:2018mtj}, the NS properties were studied using this MC-type EoS by taking $p_{\rm tr}, \Delta\varepsilon$ and $c_s$ as free parameters or by estimating them using NJL-type model. It should be noted that when $\Delta\varepsilon\to 0$, the EoS approaches to the model used in~\cite{Kanakis-Pegios:2020jnf,Kanakis-Pegios:2020kzp}. In this work, we treat these parameters as free variables to investigate their impact on GW signals.

On the other hand, if the polytropic index $\gamma \neq 0$, the EoS describes a soft phase transition termed GC~\cite{Endo:2005va,Bhattacharyya:2009fg}
\be
\varepsilon_{\rm GC}(p) = \left \{
		\begin{array}{ll}			
			\varepsilon_{h}(p),                    & p\leq p_{\rm tr}\\			
			\Lambda p^{1/\gamma_m} + p/(\gamma_m-1),     & p_{\rm tr}\leq p\leq p_{\rm css}\\			
			\varepsilon_{\rm GC}(p_{\rm css})+ (p-p_{\rm css})c^{-2}_{\rm s},                                 & p\geq p_{\rm css}			
		\end{array}		
		\right.	,
\label{eq:EoSGC}
\ee
where $\varepsilon_{\rm GC}$ stands for the energy density of the GC, $p_{\rm css}$ denotes the pressure saturating the CSS where the quark phase commences,  and unlike MC, there is no discontinuities in energy density. The constant $\Lambda$ and the polytropic index $\gamma_m$ are constrained by ensuring the continuous of the EoS at the transition points $p_{\rm tr}$ and $p_{\rm css}$. They  can be expressed equally through the energy gap $\Delta\varepsilon=\varepsilon(p_{\rm css})-\varepsilon(p_{\rm tr})$ between the beginning and ending of co-existing phase. Hence, once the CSS $c_s$ is chosen as will be done in the following, in GC, three free parameters can be identified: $p_{\rm tr}, p_{\rm css}$ and $\Delta\varepsilon$. The differences of the NS properties obtained from the MC-type and GC-type EoSs were investigated in Ref.~\cite{Bhattacharyya:2009fg} with the quark matter estimated using a bag model. Moreover, the GW signals were simulated using the GC-type EoS~\cite{Bauswein:2018bma,Bauswein:2020ggy} with some specific choices of the parameters.

In addition to the MC and GC transitions discussed above which are based on the description of a first-order phase transition between the hadron and quark phases, 
a third conceivable hadron-quark transition exists---the smooth crossover transition termed CC. 
This transition can be introduced phenomenologically through the pressure function as~\cite{Masuda:2012ed,Masuda:2012kf}
\be
p(n) & = & p_{h}(n)f_{-}(n) + p_{q}(n) f_{+}(n),
\label{eq:crossP}
\ee
where $p_h$ and $p_q$ are, respectively, the pressure function described in terms of hadrons and quarks. The hadron part pressure function $p_{h}(n)$ can be obtained from the energy density $\varepsilon_h(n)$ in the MC~\eqref{eq:EoSMC} and GC~\eqref{eq:EoSGC} by using the thermal relation~\eqref{eq:thermal}. Similarly, the quark part of the CC EoS $p_{q}(n)$ is constructed by using the CSS parameterization of $\varepsilon_{\rm MC}(n)$ for $p\ge p_{\rm tr}$ in~\eqref{eq:EoSMC} through the thermal relation~\eqref{eq:thermal}. Extrapolating $p_{h}$ and $p_{q}$ to the whole density region and joining them using a smooth function, one obtains the EoS of crossover transition. The smooth function is taken typically as 
\be
f_{\pm}(n) & = & \frac{1}{2}\left(1\pm\tanh\left(\frac{n-\bar{n}}{\Gamma}\right)\right),
\ee
where $\bar{n}$ denotes the central value of the crossover density region, and $\Gamma$ accounts for the intensity of the transition. In the context of the crossover transition, the EoS comprising hadronic matter and quark matter forms a globally continuous function of density, blurring the distinction between the hadron and quark phases. The EoS of the crossover transition constructed in such a way has parameters $p_{\rm tr}, \Delta\varepsilon, \bar{n}$ and $\Gamma$. These parameters can be estimated by requiring that the EoS of CC continuously and monotonously converges to that of MC at low and high densities. A detailed discussion of the difference of the above three types of hadron-quark transitions with possible parameter choices and their effects on GW signals will be given later. The star properties were explored with or without strangeness by varying the intensity of the transition in Refs.~\cite{Masuda:2012ed,Masuda:2012kf,Kapusta:2021ney} with quark matter calculated using the Nambu–Jona-Lasinio (NJL) model. Here, the EoS of quark matter is parameterized in terms of the CSS approximation.


\section{Methodology}

\label{sec:metho}

In this section, for completeness, we summarize the main points of the numerical simulation used in this work. The details were presented in the referred literature.  

For the purpose to illustrate the impacts of varying constructions of the hadron-quark transition and associated parameters, in the low densities of compact star matter, we choose pure hadronic $\varepsilon_h(p)$ from the typical pure-nucleon APR EoS~\cite{Akmal:1998cf}. Furthermore, in the case of MC and GC, following~\cite{Montana:2018bkb}, we take the CSS parameterization of the EoS which can be viewed as the lowest-order terms of a Taylor expansion of the high-density EoS around the transition pressure~\cite{Zdunik:2012dj,Alford:2013aca}. At lower density after $p_{\rm tr}$ in MC and after $p_{\rm ccs}$ in GC, we take $c^2_s=1$ to offer a stiff EoS. Conversely, at higher density, the value is taken to be $c^2_s=1/3$ to approach the conformal limit in perturbative QCD. A comparison of the EoSs for these three constructions will be depicted in the right panel of Figure~\ref{fig:35-150}. 

In the numerical simulation of the GW signals, we apply the following nine-segment isentropic polytropic approximation of the aforementioned EoS for the entire density region--- both the low-density region of hadronic matter and the high-density region of quark matter---of the neutron star matter up to $\sim 8.0 n_0$,
\be
p_i & = & K_{i}n^{\Gamma_{i}}, \nonumber\\
\varepsilon_i & = & \varepsilon_{i-1}n_{i-1}+\frac{K_{i}}{\Gamma_{i}-1}n^{\Gamma},
\label{eq:EoSpiece}
\ee
where $\varepsilon_{i-1}$ and $n_{i-1}$ $(i=1-9)$ are the energy density and baryon number density of the $i-1$ segment end point, 
$K_{i}$ and $\Gamma_{i}$ are the polytropic constant and index of the $i$-th segment, respectively. Explicitly, the segmentation is obtained by optimizing the density location for each segmentation with $K_i$ and $\Gamma_i$ fitted such that the minimal deviation between the fitted EoS and original EoS is arrived at. For different constructions, the transition points $p_{\rm tr}$ and $p_{\rm css}$ are taken to be the same in parameterzing the Maxwell and Gibbs constructed EoSs in order to insure that the only difference between EoSs is from the types of constructions. Note that we artificially increased the sound speed slightly, i.e., $c_s^2 = dp/d\varepsilon = 10^{-4}$, during the phase transition of the MC to avoid possible numerical errors that leads to unphysical results.

Furthermore, we also take the thermal correction into account through an ideal, nonrelativistic fermion gas approximation~\cite{Keil:1995hw}. In this approximation, the pressure used in the GW simulation is decomposed as
	\be
	p & = & p_{\rm cold}(n) + p_{\rm th}, 
	\label{eq:EoSFull}
	\ee 
	where $p_{\rm cold}$ is the EoS at zero temperature discussed above, $p_{\rm th}$ is the thermal correction of the pressure. In the ideal, nonrelativistic fermion gas approximation, 
	\be
	p_{\rm th} & = & \left( \Gamma_{\rm th} -1\right)n \varepsilon_{\rm th}
	\ee
	where $\varepsilon_{\rm th}$ is the thermal part of specific internal energy, and
	thermal adiabatic index $\Gamma_{\rm th}$ can be taken as $\Gamma_{\rm th} \simeq 1.75$~\cite{Bauswein:2010dn}.

The GWs emitted from BNS mergers were investigated by performing the numerical relativity simulations by using the Einstein Toolkit~\cite{Loffler:2011ay,Zilhao:2013hia}, which is a community-driven software
platform of core computational tools~\cite{Schnetter:2003rb,Baiotti:2004wn,Hawke:2005zw,Mosta:2013gwu}. Generally, the simulation can be decomposed into two steps. The first step is to generate the initial data~\cite{Read:2008iy}, such as the initial frequency and angular momentum of the equilibrium state of the BNS system before merge, by using the initial conditions~\cite{Gourgoulhon:2000nn} of irrotational binary systems, i.e., the separation distance, component masses and the EoS~\eqref{eq:EoSpiece}. This step is fulfilled by using the public codes of the LORENE library (https://lorene.obspm.fr/). The second step is the relativistic hydro-simulation which is performed utilizing the Einstein Toolkit~\cite{Loffler:2011ay,Zilhao:2013hia}, a numerical relativity simulation platform based on the Cactus Computational Toolkit~\cite{Schnetter:2003rb,Baiotti:2004wn,Hawke:2005zw,Mosta:2013gwu}(https://www.cct.lsu.edu/$\sim$eschnett/McLachlan/; http://cactuscode.org/)---a software framework for high-performance computing to advance and support research in relativistic astrophysics and gravitational physics. In this step, the initial date generated by the first step, the grid resolution and the parameterized EoS including the thermal contribution~\eqref{eq:EoSFull} are input. The main tools that we used are Carpet-Adaptive mesh refinement methods, GRhydro-Hydrodynamic evolution, and McLachlan-spacetime metric evolution. We used a fourth-order Runge-Kutta method for solving the differential equation and fourth-order finite difference stencils for spacetime evolution, HLLE for Riemann solver, weno for reconstruction, and the system grid uses a $2$D symmetry across the $x$-$y$ plane. The grid resolution in our simulation is taken as $\sim260$~m.

Using the Einstein Toolkit module WeylScal4, the GW signal is extracted from each simulation with the standard method of calculating during the numerical evolution of the Newman–Penrose scalar $\psi_4$~\cite{Maione:2016zqz}. Its asymptotic limit (for $r \to \infty$) describes gravitational radiation
\be
\psi_4 = \ddot{h}_+ + i \ddot{h}_\times = \ddot{\bar{h}}
\ee
where $h$ is the complex GW strain $h=h_++ih_\times$ with $h_+$ and $h_\times$ being the polarizations in the transverse–traceless (TT) gauge. The Newman–Penrose scalar $\psi_4$ is computed in a spherical surface far away from the source, and then decomposed in spin-weighted spherical harmonics of weight-2
\be
\psi_4(t,r,\theta,\phi) & = & \sum_{l=2}^{\infty} \sum_{m=-l}^{l} \psi_4^{lm}(t,r)Y_{lm}(\theta,\phi) 
\ee
which is done by the thorn Multipole. Since $\psi_4$ falls off as $1/r$ in asymptotically flat space, often the results for $r\psi_4$ (and equivalently $rh$) are reported. To get the GW strain $h^{lm}$, the double time integration should be numerically done by using a simple trapezoidal rule~\cite{Damour:2009wj} formalized in a first-order gauge-invariant representation of the variables~\cite{Abrahams:1990jf,Anninos:1994vw,Abrahams:1995gn} as
\be
\bar{h}^{(0)}_{lm}(t,r) & = & \int_0^tdt^\prime\int_0^{t^\prime} dt^{\prime\prime}\psi_4^{lm}(t^{\prime\prime},r),\nonumber\\
\bar{h}_{lm}(t,r) & = & \bar{h}_{lm}^{(0)} - Q_1 t -Q_0
\ee
where 
\be
Q_1 & = & \frac{\partial \bar{h}}{\partial t}|_{t=0}, ~~~~ Q_0 = \bar{h}(t=0)
\ee
being the integration constants fitting to the resulting strain with first order polynomial and to be subtracted from the strain itself. To eliminate the low-frequency unphysical oscillations in the strain amplitude caused by unresolved high-frequency noise aliased in the low-frequency signal during the integration
process, we further subtract from the strain a second order polynomial fit~\cite{DePietri:2015lya}
\be
\bar{h}_{lm}(t,r) & = & \bar{h}_{lm}^{(0)} - Q_2 t^2 - Q_1 t -Q_0
\ee
which are found sufficient to eliminate the unphysical drift from the dominant $(2,2)$ mode considered in this work. In the following, instead of $\bar{h}_{lm}(t,r)$, we will present the results of $r\bar{h}_{22}(t-r_{\rm ret},r)$ which are constant of $r$ when $r \to \infty$ with $r$ and $t_{\rm ret}$ being, respectively, the radius and retarded time at which the GW signal is extracted. Consequently, the GW signal in time-domain, as illustrated in the following figures, can be extracted.

The time-frequency-amplitude
relation can be obtained by using the short-time Fourier
transform (STFT)~\cite{allen:1977unified,sejdic2009time,allen1977short}
\be
\tilde{h}^{lm}(t,f) & = & \int_{-\infty}^{\infty} h^{lm}(t^\prime) w(t-t^\prime) e^{-2\pi i f(t-t^\prime)} dt^\prime
\ee
where $w(t)$ is the window function.

In addition, after obtaining the $h^{lm}(t)$ with the above procedure, the radiated energy can be obtained via
\be
\frac{dE}{dt} & = & \frac{R^2}{16\pi}\int d\Omega |\dot{h}(t,\theta,\phi)|^2.
\ee


\section{Impact of the parameters in the EoS on GW signals}

\label{sec:GWTr}

We first discuss the impact of the parameters in different constructions on the GWs of BNS mergers. For this purpose, we choose the equal mass systems with component mass $1.1M_\odot$, with an initial separation of $40$~km.  

\begin{figure}[htb]
	\includegraphics[width=0.5\linewidth]{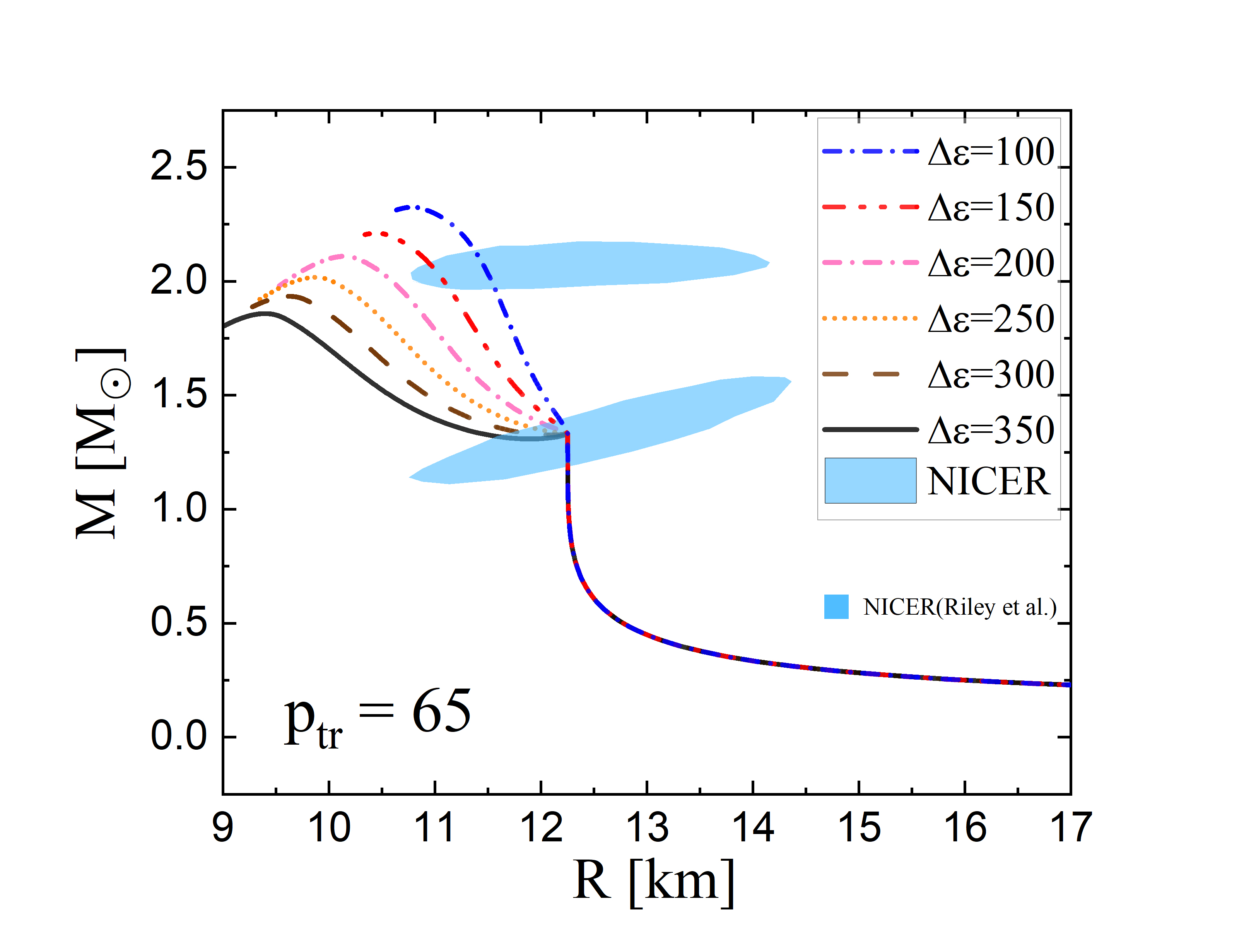} 
	\includegraphics[width=0.5\linewidth]{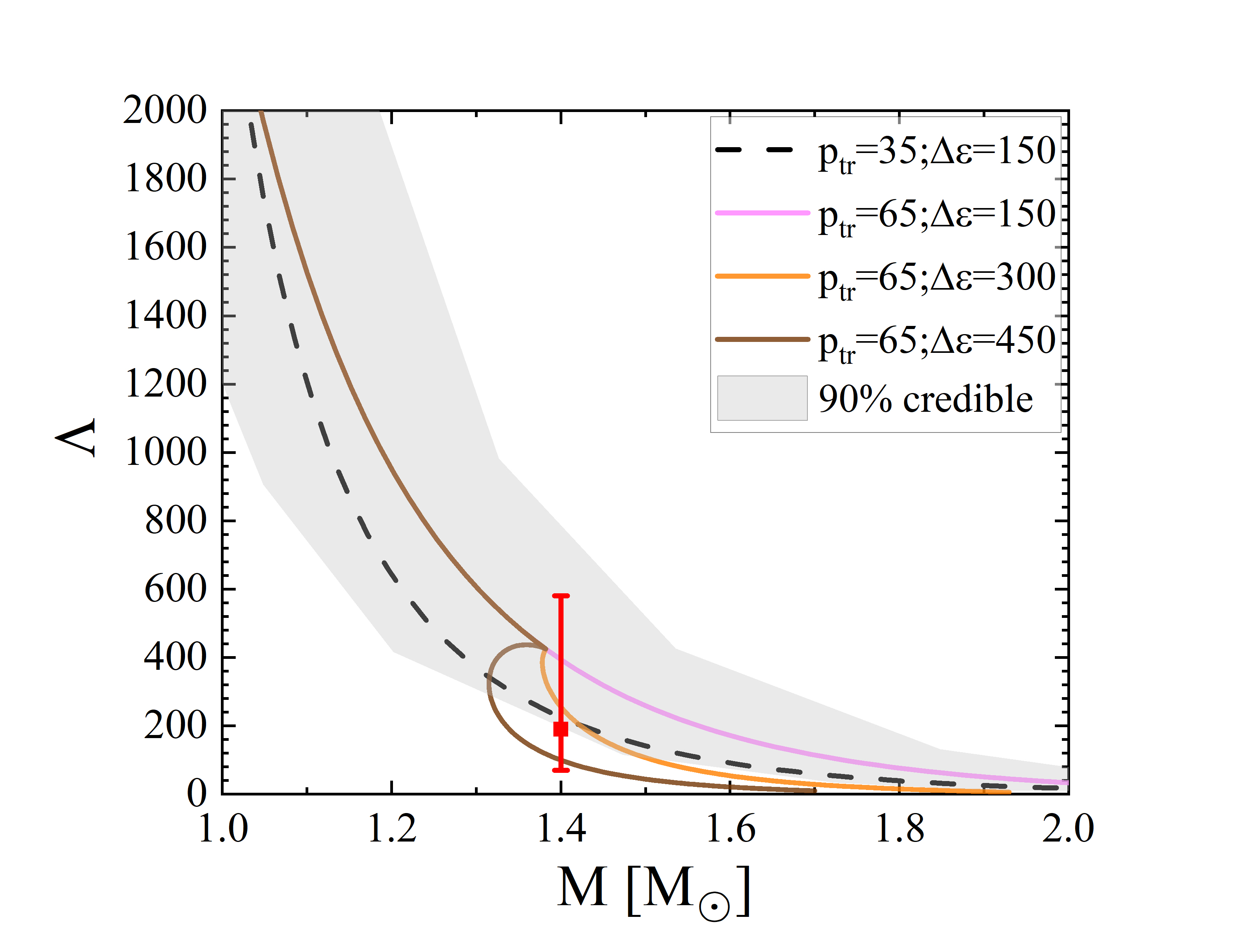}
	\caption{Mass-radius relation (left panel) and tidal deformability (right panel) calculated using the MC-type EoS. The constraints in the left panel (the light-blue area) are taken from~\cite{Riley:2019yda,Riley:2021pdl}. In the right panel, the red-vertical bar is the tidal deformability for the NS with $1.4M_{\odot}$ extracted from GW170818 $\Lambda_{1.4M_{\odot}}=190^{+390}_{-120}$~\cite{Tsang:2018kqj} and the gray constraint is from Ref.~\cite{Ferreira:2021osk}. Both $p_{\rm tr}$ and $\Delta \varepsilon$ are in units of MeV/fm$^{3}$.}
	\label{fig:starMC}
\end{figure}

We depict the global properties of NSs calculated using MC-type transition in Figure~\ref{fig:starMC}. The left panel indicates that the smaller latent heat $\Delta\varepsilon$, the stiffer the EoS after phase transition. When the nuclear matter at low density is described by the APR EoS, the latent heat is estimated to be $\lesssim 200$~MeV/fm$^3$ with respect to the constraint imposed by NICER. The right panel tells us that the tidal deformability calculated from the current parameter choice saturates the constraints extracted from GW170817.

From Figure~\ref{fig:starMC} one can see that a quark core appears in NSs with mass $\gsim 1.25 M_{\odot}$ when $p_{\rm tr}=65$~MeV/fm$^{3}$ ($\sim 2.8 n_0$) and the NSs become hybrid ones. When the latent is large, e.g., $\Delta\varepsilon \gsim 350$MeV/fm$^3$, the twin star scenario arises due to the hadron-quark phase transition~\cite{Glendenning:1998ag,Alford:2017qgh,Alvarez-Castillo:2017qki}. Since the initial configurations we considered are either the BNS systems with mass $1.1 M_{\odot}$ or heavier star with $\Delta\varepsilon = 150$MeV/fm$^3$, the twin star configuration does not affect initial choice.

\begin{center}
\begin{figure}[htb]
	\includegraphics[width=1.0\linewidth]{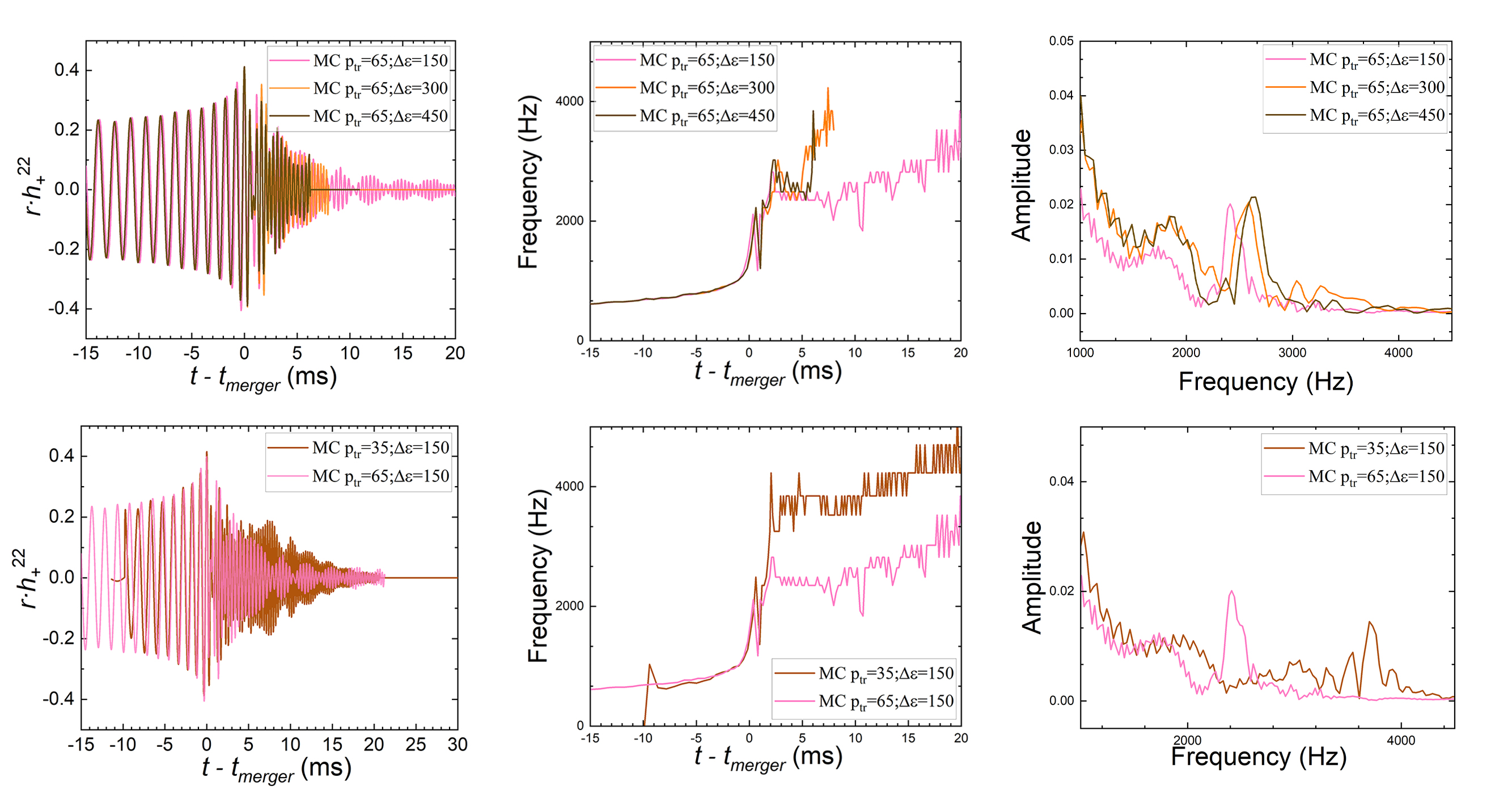}
	\caption{Gravitational wave signals from the BNS mergers with MC of transition. Both $p_{\rm tr}$ and $\Delta \varepsilon$ are in units of MeV/fm$^{3}$.}
	\label{fig:GWMC}
\end{figure}
\end{center}

The GW signals resulting from the BNS mergers using the MC-type transition are plotted in Figure~\ref{fig:GWMC}. In the upper row we choose $p_{\rm tr} = 65$~MeV/fm$^3$ ($\sim 2.8 n_0$) and vary latent energy $\Delta\varepsilon$ while in the lower row we choose $\Delta\varepsilon=150$~MeV/fm$^3$ and vary $p_{\rm tr}$ for the purpose to investigate the model-parameter dependence of the GW signals. The upper-left panel depicts the waveforms obtained with $p_{\rm tr} = 65$~MeV/fm$^3$, albeit varying latent energy $\Delta\varepsilon$. We can see that the differences between the waveforms start to appear only after the merger of the binary stars. This is because, for the star with mass $1.1M_\odot$, the hadron-quark transition does not affect the star properties, so the GWs in the inspiral phase are not affected by the transition. However, in the post-merger phase, since the central density of the remnant is bigger than the density where the transition occurs, the waveforms are notably affected by the hadron-quark transition. The upper-middle panel illustrates the evolution of the GW frequencies. One can see that the frequencies remain nearly constant during the inspiral phase since the two stars are far away from each other and frequency of the inspiral is nearly steady. However, when the two stars are close in the inspiral phase and in the merger phase, the frequency steadily rises due to the violent matter exchange. A noticeable trend emerges where a higher latent heat, or softer EoS, leads to a swifter frequency increase. The frequency spectrum is shown in the upper-right panel. The result shows that, after chirp---the coalescence the two stars---the frequency spectrum keeps decreasing. Given the phase transition density exceeds the core density of the parent stars, the chirp frequencies and spectrum are not different so much.

%
\begin{figure}[phtb]
	\includegraphics[width=1\linewidth]{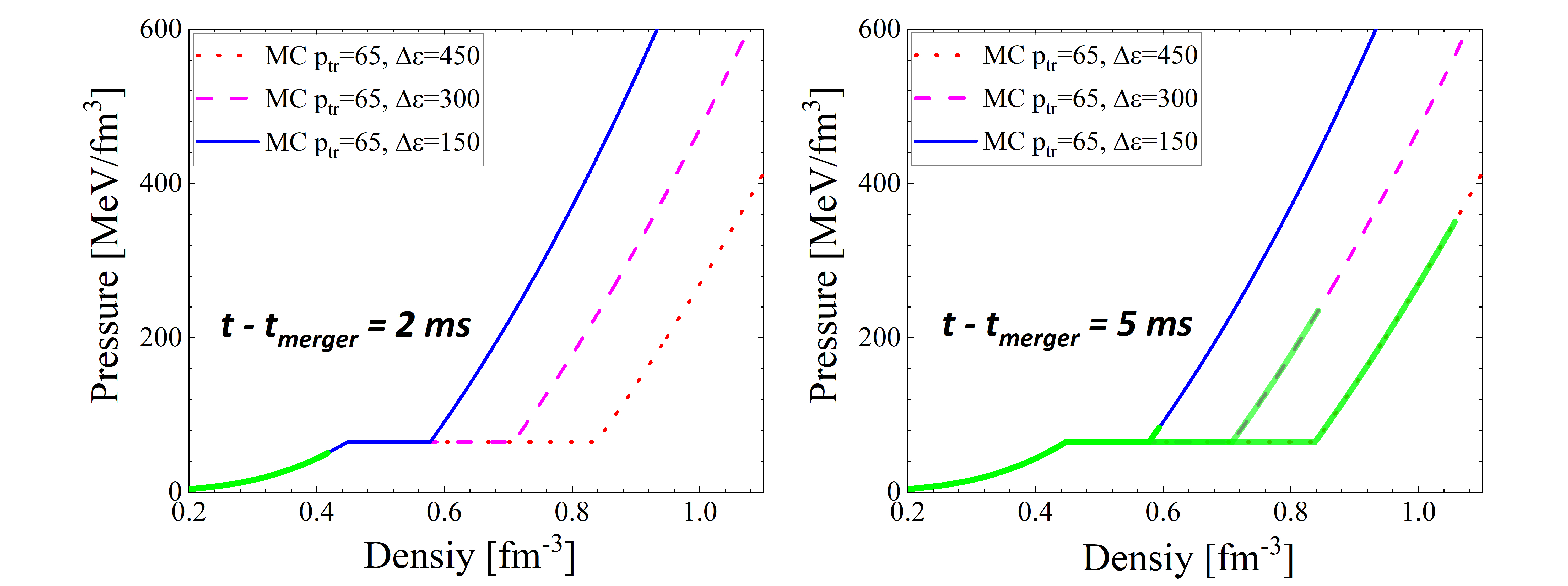}
	\caption{Evolution of the EoSs covered in the BNS merger (green lines) in the left panel of Figure~\ref{fig:GWMC}. Both $p_{\rm tr}$ and $\Delta \varepsilon$ are in units of MeV/fm$^{3}$.}
	\label{fig:maxrho}
\end{figure}
%

The green line in Figure~\ref{fig:maxrho} shows the evolution of the EoSs covered in the BNS merger. One can clearly see that, up to the post-merger time $t-t_{\rm merger}=2$~ms, the phase transition does not enter the evolution so that the GWs are the same. However, at $t-t_{\rm merger}=5$~ms, the central densities are in the quark phase and the bigger the latent heat, the larger the cores made of quark involved and, therefore the GWs are different in the post-merger phase.

After merger, the remnant of BNS can form a differential rotating hyper-massive neutron star (HMNS) exceeding the maximum mass derived from TOV equations. With the decreasing of differential rotation, the HMNS would collapse to a black hole (BH), indicated by the null amplitude of the GWs shown in the figure. The duration of the ring-down period depends on the EoS. As one can see from the upper row of Figure~\ref{fig:GWMC}, the bigger the latent heat $\Delta\varepsilon$, the shorter HMNS duration before the formation of BH, that is, the shorter the ring-down period and, the bigger the frequency after merger. These phenomena can be elucidated by the softness of the EoS attributed to the phase transition.

The lower row of Figure~\ref{fig:GWMC} showcases the GW signals stemming from the BNS mergers with $\Delta\varepsilon=150$~MeV/fm$^3$ but different values of $p_{\rm tr}=35$~MeV/fm$^{3}$ ($\sim 2.2 n_0$) for the purpose to elucidate the effect of the location of the transition. Different from the upper-left panel, the GW waveforms in this scenario exhibit substantial discrepancies even in the inspiral phase. With a lower $p_{\rm tr}$, i.e., a larger quark core exists in the NS, the binary system would merge quicklier compared to those with a higher $p_{\rm tr}$. Therefore, after merger, the frequency of the GW calculated with $p_{\rm tr}=65$~MeV/fm$^3$ is smaller than that calculated with $p_{\rm tr}=35$~MeV/fm$^3$ (lower-middle panel) and the chirp frequency spectrum is smaller for $p_{\rm tr}=65$~MeV/fm$^3$ (lower-right panel).

The variation of the GW signals in the inspiral phase can be attributed to the tidal deformability $\Lambda$ of the NSs involved, which describes the deformability or the level of difficulty in matter exchange in the BNS system. As illustrated in the right panel of Figure~\ref{fig:starMC}, for the selected parameters from the lower row of Figure~\ref{fig:GWMC}, $\Lambda\simeq1540$ for $p_{\rm tr}=65$~MeV/fm$^3$ and $\Lambda\simeq1160$ for $p_{\rm tr}=35$~MeV/fm$^3$. Consequently, the duration of the inspiral phase is shorter for the latter scenario.

\begin{center}
		\begin{figure}[tbp]
			\includegraphics[width=1\linewidth]{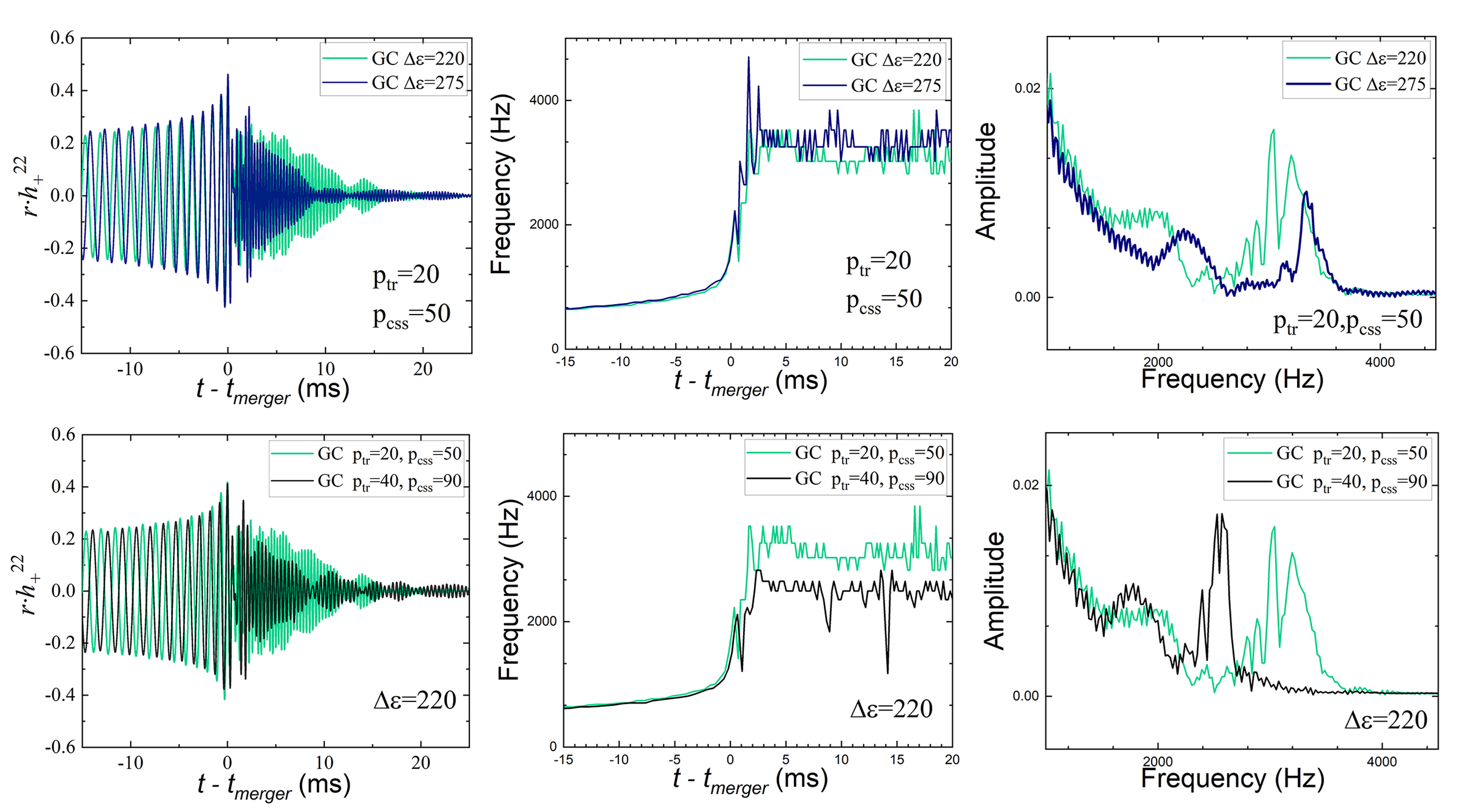}
			\caption{Gravitational waves from BNS mergers with GC-type transition. Units of $p_{\rm tr}$, $p_{\rm css}$ and $\Delta \varepsilon$ are MeV/fm$^{3}$.
			}
			\label{fig:gc11vs11}
		\end{figure}
	\end{center}

We next show the GW signals from BNS mergers simulated with GC of the transition in Figure~\ref{fig:gc11vs11}. In the upper row, the GWs from BNS mergers with the same transition pressure range ($p_{\rm tr}$ to $p_{\rm css}$) but varying $\Delta\varepsilon$ are displayed, while the lower row shows the simulations with the same $\Delta\varepsilon$ but diverse pressure range. In both left panels, the GWs exhibit differences in both the inspiral and post-merger phases because the hadron-quark transition enters the core of the star with mass $1.1M_{\odot}$ and the difference of the latent heat effects. The upper-left panel demonstrates that with a smaller $\Delta\varepsilon$, the frequency is smaller, albeit marginally, while the amplitudes in the post-merger phase is bigger. Conversely, the lower-left panel indicates that a narrower range of the transition pressure leads to a slightly higher inspiral frequency and bigger GW amplitudes in the post-merger phase. From the middle column, one can see that the frequency remains roughly constant in the post-merger phase, slightly decreases with latent heat (upper panel) while decreases with location of the phase transition pressure (lower panel). The right column tells us the following information: a higher latent heat value results in the peak of frequency spectrum shifting toward a higher value while its magnitude is diminished. Conversely, when the phase transition location is reduced, the peak of frequency spectrum is shifted toward a higher value.

\begin{center}
\begin{figure}[hptb]
	\includegraphics[width=1.0\linewidth]{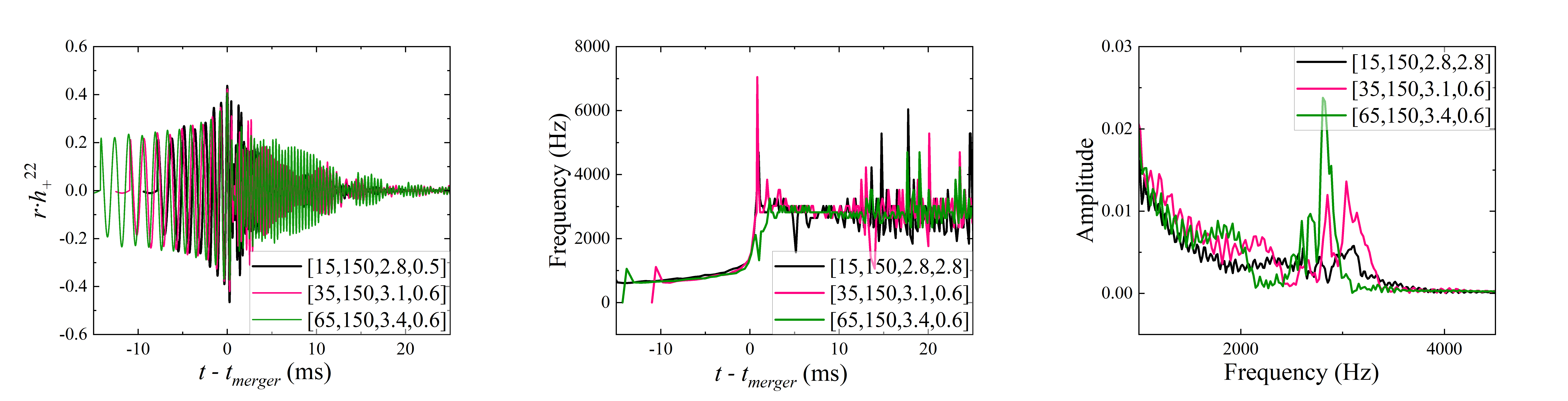}
	\caption{GW signals from BNS mergers with CC-type transition. In the parameter set $[p_{\rm tr}, \Delta\varepsilon, \bar{n}, \Gamma]$, $p_{\rm tr}$ and $\Delta\varepsilon$ are in units of MeV/fm$^{3}$ and, $\bar{n}$ and $\Gamma$ are dimensionless according to our convention.}
	\label{fig:cross11vs11}
\end{figure}
\end{center}

Finally, in Figure~\ref{fig:cross11vs11}, we present the GW signals resulting from the CC type transition. To investigate the impact of the transition region, we select three sets of parameters $[p_{\rm tr}, \Delta\varepsilon, \bar{n}, \Gamma]$ as shown in the figure, ensuring the coincidence of the EoSs at low and high densities. It is found that the NS properties from all these three sets of parameters are consistent with NICER~\cite{Riley:2019yda,Riley:2021pdl}. It clearly shows that a lower transition density leads to a shorter inspiral phase, a briefer post-merger phase, and reduced wave amplitudes in the post-merger phase. The evolution of frequency (middle panel) indicates that in both the inspiral and post-merger phases it is not changed so much. In contrast, the magnitude of the peak of frequency is sensitive to the parameters involved and therefore suggests a possible means to distinguish the EoS of CC.

\section{Effect of component masses}

\label{sec:CM}

We next discuss the variation of the GWs arising from the selection of component masses of the BNS system.

\begin{figure}[htbp]
	\includegraphics[width=1\linewidth]{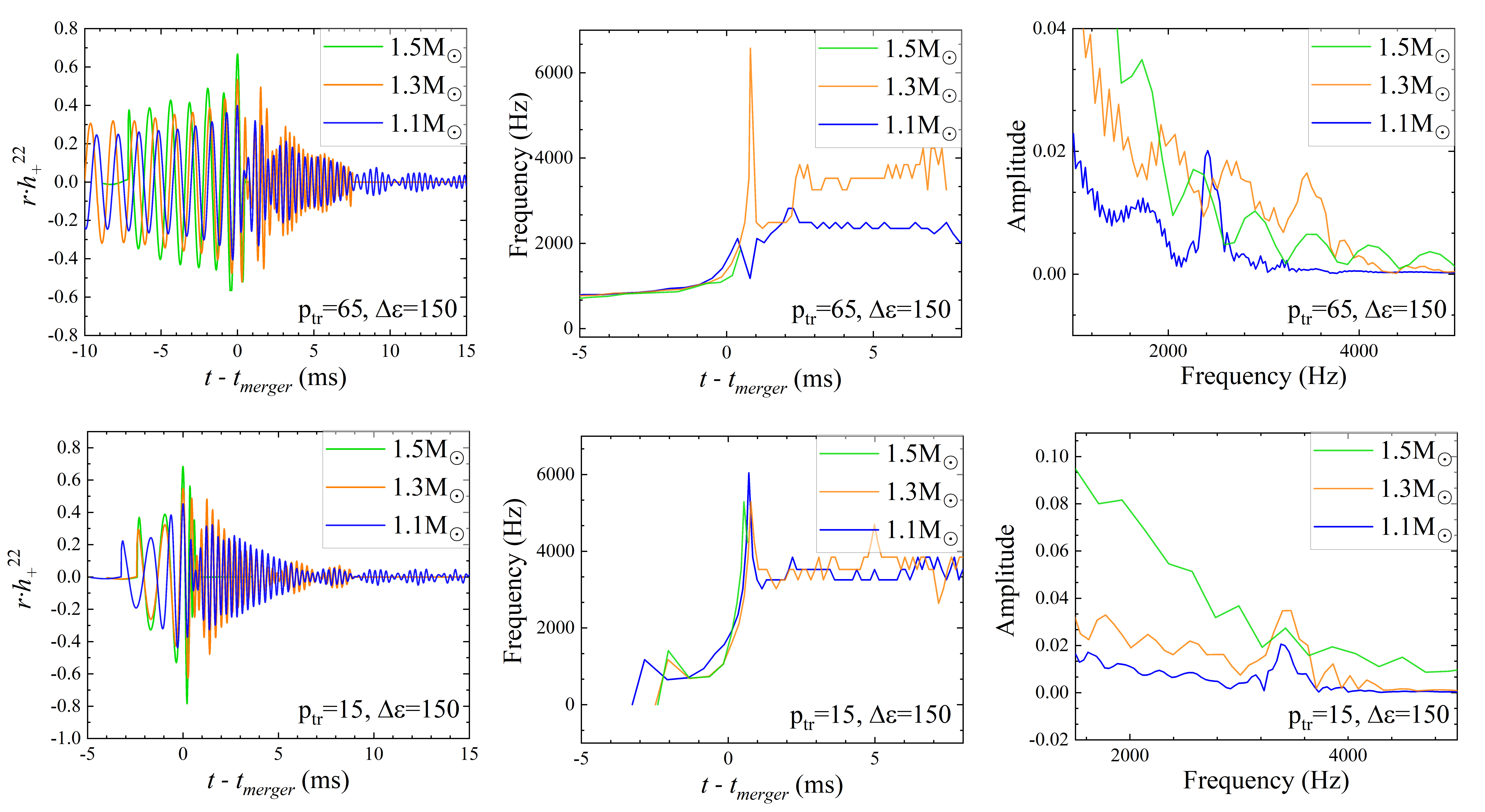}
	\caption{Gravitational waves from the BNS mergers with MC-type transition with different component masses. Both $p_{\rm tr}$ and $\Delta \varepsilon$ are in unit of MeV/fm$^{3}$.}
	\label{fig:mcmass}
\end{figure}

We first set $p_{\rm tr} = 65$~MeV/fm$^3$ and $\Delta \varepsilon=150$~MeV/fm$^3$ and consider the equal mass system with component masses $1.1M_{\odot}, 1.3M_{\odot}$ and $1.5M_{\odot}$. From the left panel of Figure~\ref{fig:starMC} one concludes that the cores of stars with masses  $1.3M_{\odot}$ and $1.5M_{\odot}$ are in the quark phase although their sizes are different. The upper row of Figure~\ref{fig:mcmass} clearly demonstrates that the component mass significantly impact the maximum amplitude of GW and merger time of BNS. As the mass of the individual star increases, the inspiral period shortens, the maximum amplitude grows, and the ring-down period decreases. These distinctions align with our previous discussions on the stiffness of the EoS. The frequency spectrum shown in the upper-middle panel indicates that when the quark matter enters the cores of the parent stars, the frequency becomes larger compared to that parent stars without quark cores. One can conclude from the upper-right panel that at low frequency the amplitude of GW stemming from the BNS with heavier component mass is easier to detect but at high frequency the amplitude is not sensitive to the component mass.

To investigate whether the above conclusion on the effect of component mass depends on the location of phase transition, we illustrate the results from $p_{\rm tr}= 15$~MeV/fm$^{3}$ but fix other parameters in the lower-row of Figure~\ref{fig:mcmass}. It can be easily seen that the inspiral period is further shorten due to the larger quark core (lower-left panel), the frequency domain is not sensitive to the component mass since the three combinations considered all include quark core (lower-middle panel) and, the tendency of the frequency dependence of the amplitude is intact (lower-right panel).

The same conclusions and reasoning apply to the results from the GC- and CC-type transitions so we will not repeat them here.

\section{Effect of the type of the hadron-quark transition}

\label{sec:Type}

\begin{figure}[htb]
	\centering
	\includegraphics[width=1\linewidth]{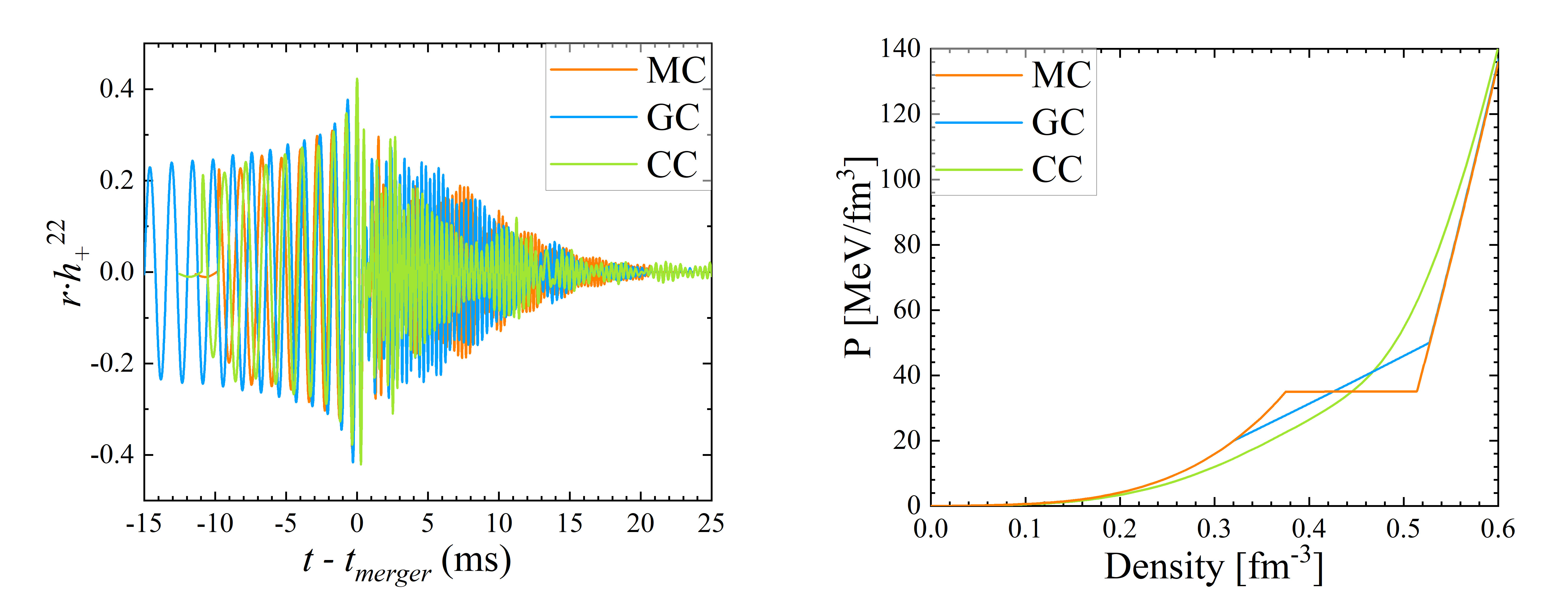}
	\caption{Gravitational waves from the BNS mergers with different constructions of the hadron-quark transition.}
	\label{fig:35-150}
\end{figure}

Finally, we shall discuss how the nature of the hadron-quark transition impacts the GWs arising from the BNS mergers. For this purpose, we choose the typical values of the parameters of MC as $(p_{\rm tr}, \Delta\varepsilon)=(35,150)$~\mbox{MeV/fm}$^3$ and $(p_{\rm tr},p_{\rm css}, \Delta\varepsilon)=(20, 50, 223)$~\mbox{MeV/fm}$^3$ for the GC. With these parameters, we ensure a precise alignment of the EoS of MC and GC pre and post the phase transitions, as shown in the right panel of Figure~\ref{fig:35-150}. Similarly, as shown in Figure~\ref{fig:35-150}, the CC is constructed to ensure that the EoS is monotonically and continuously convergent to that of MC at low and high densities. With this requirement, the parameter set is taken to be $[p_{\rm tr}, \Delta\varepsilon, \bar{n}, \Gamma]= [35~\mbox{MeV/fm}^3, 150~\mbox{MeV/fm}^3, 3.1, 0.6]$ In such a way, the differences in the GWs are only from the constructions of the transition.

In the left panel of Figure~\ref{fig:35-150}, we present the simulations for a binary neutron star with equal masses of $1.1M_{\odot}$. One can see that in the inspiral phase, the amplitudes of the GWs arising from the MC and CC are akin but smaller than those from the GC. Among the three constructions, the MC results in the shortest inspiral period, while the GC yields the longest inspiral period. With the increasing maximum density and pressure during the merger process, the transition starts to play a considerable role in the post-merger phase. This is reflected in the differences of waveforms after $t-t_{\rm merger}>0$~ms, as shown in the figure, that the CC gives a much smaller amplitude than that from MC and GC.

\begin{center}
\begin{figure}[htbp]
	\includegraphics[width=1\linewidth]{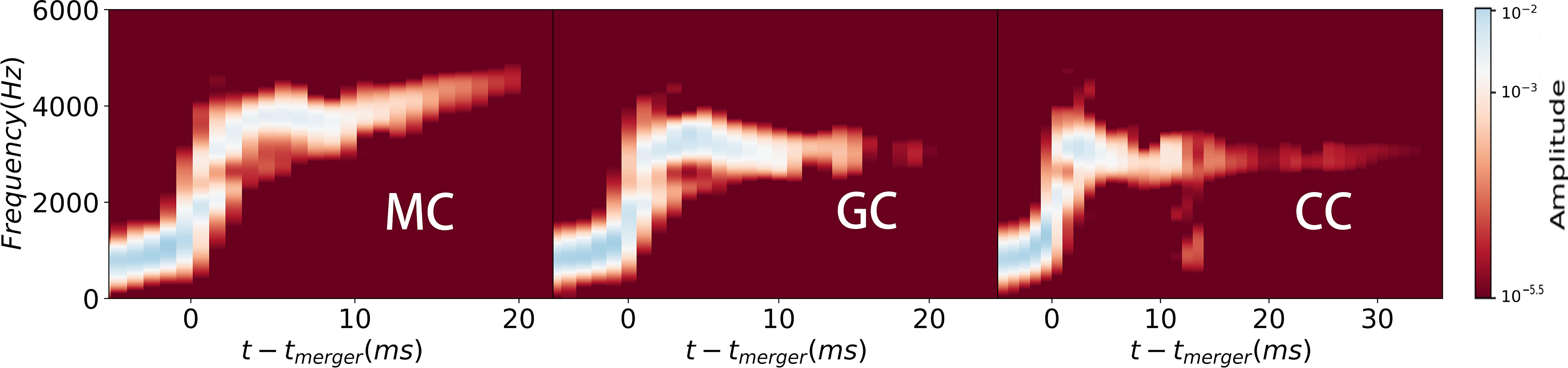}
	\caption{Time-frequency of GW from the BNS mergers with various constructions of hadron-quark transition. The amplitude of gravitational waves is represented by the brightness of the colors. The parameter choice is the same as that in Figure~\ref{fig:35-150}.}
	\label{fig:fft}
\end{figure}
\end{center}

The impact of transition becomes more evident in the time-frequency diagram in Figure~\ref{fig:fft}, derived from  Figure~\ref{fig:35-150} using the short-time Fourier transform technique. Across all the three constructions, the GW frequencies showcase a swift rise during the inspiral phase at $t-t_{\rm merger}< 0$, accompanied by peak GW amplitudes indicated by the brighter colors in the graphs. Notably, a compelling characteristic is that the GW frequency from the Maxwell construction increases even after the merger while those from the Gibbs and crossover constructions decrease.

\begin{figure}[htb]
	\centering
	\includegraphics[width=0.45\linewidth]{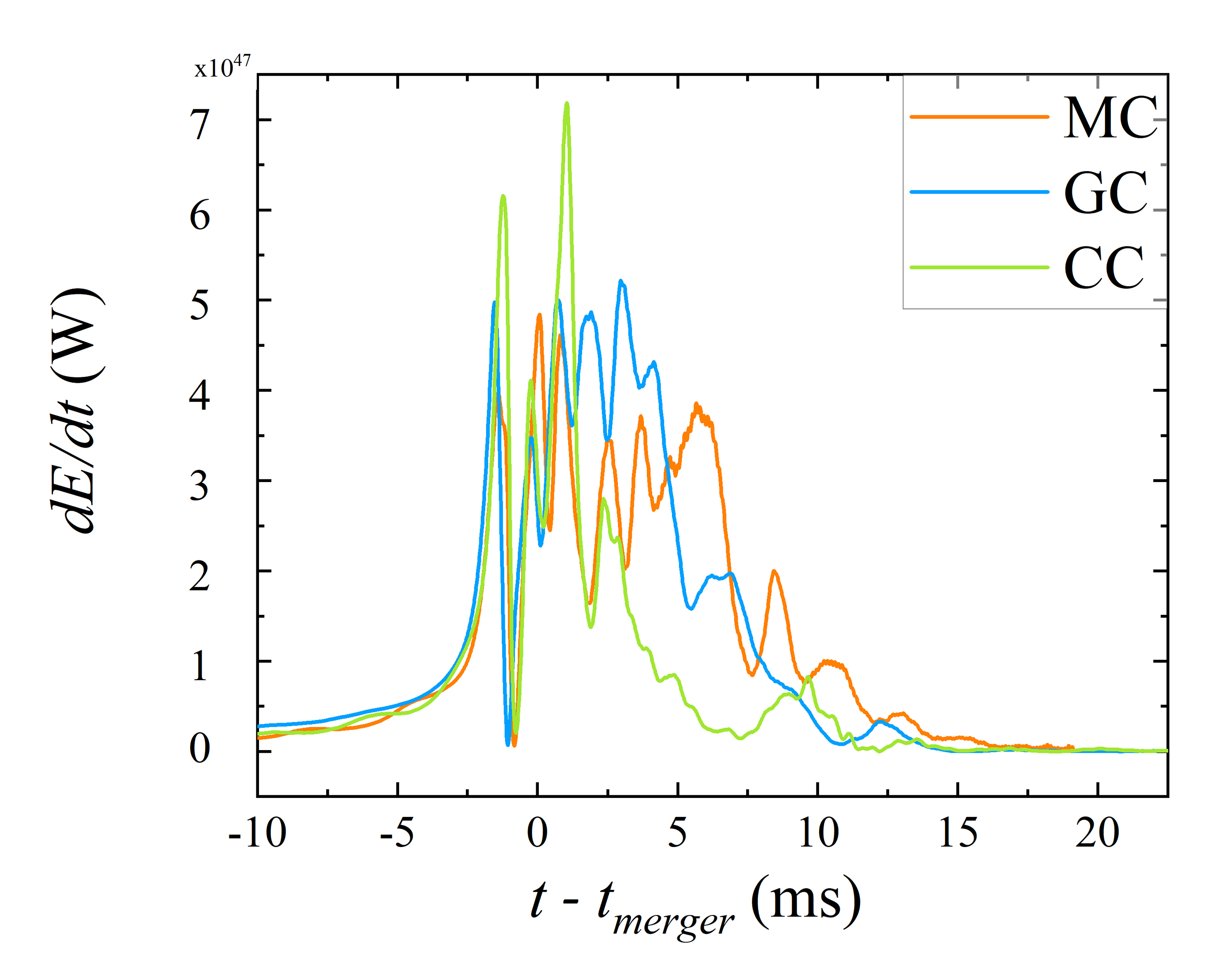}	\includegraphics[width=0.48\linewidth]{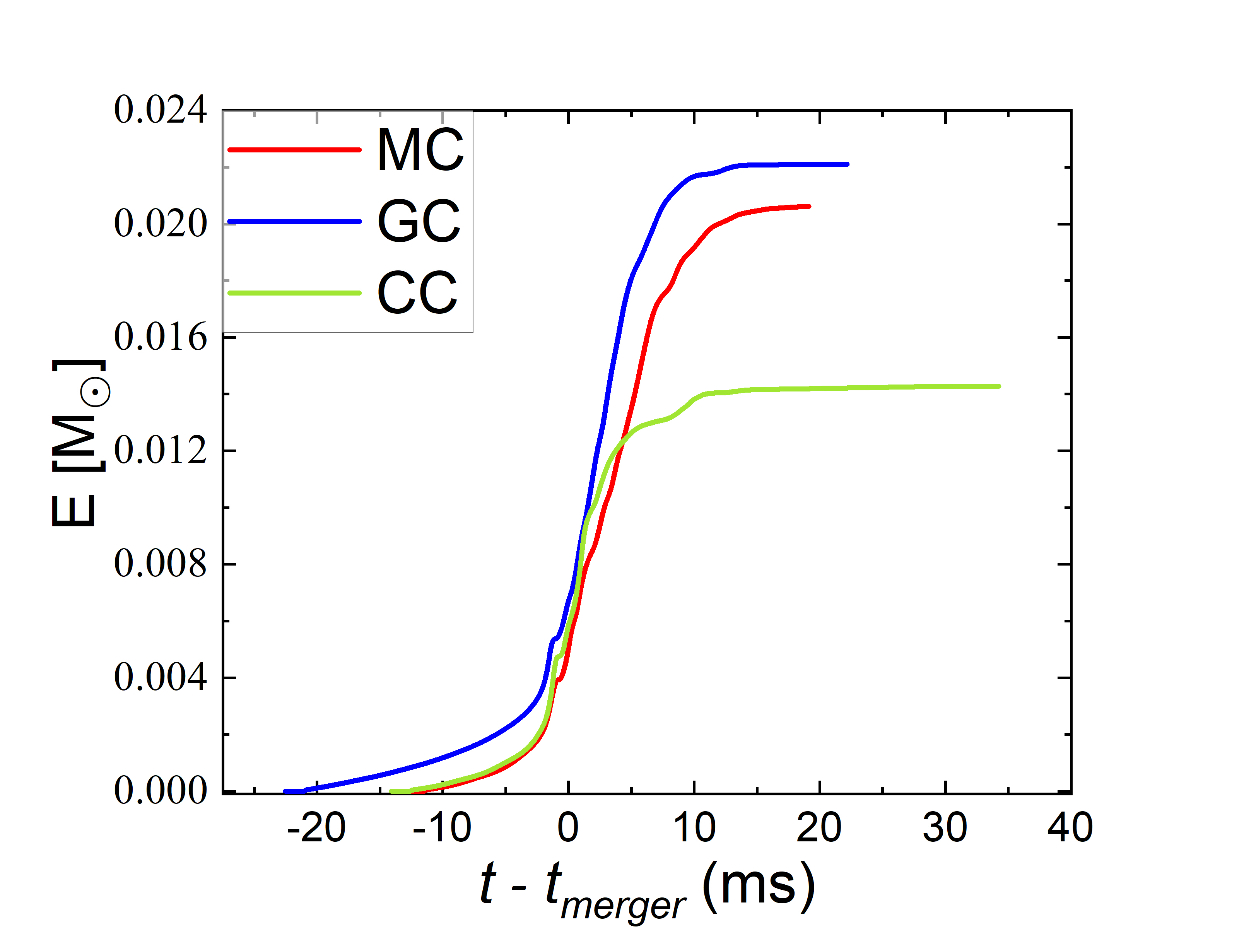}
	\caption{GW luminosity (left panel) and radiated energy (right panel) of BNS systems displayed in Figure~\ref{fig:35-150}.}
	\label{fig:luminosity}
\end{figure}

We compare the radiated energy (power) carried by gravitational waves with different constructions of the transition in Figure~\ref{fig:luminosity}. The EoSs we use are the same as those utilized in Figure~\ref{fig:35-150}. One can see that, the widths of the powers vary depending on the specific EoSs, the bigger the latent heat, the broader the power for MC and GC. Since hadronic and quark components coexist across all the density ranges within the NSs, the power spectrum is narrower for the CC scenario. Regarding the radiated energy, one can estimate the remnant masses as $\Delta m \approx (2.18,2.178,2.186)M_\odot$ for, respectively, the MC, GC, and CC. Concerning the maximum NS masses available within these three constructions, one can easily understand the waveforms of Figs.~\ref{fig:GWMC},~\ref{fig:gc11vs11} and \ref{fig:cross11vs11}.

\begin{center}
\begin{figure}[htb]
	\centering
	\includegraphics[width=1\linewidth]{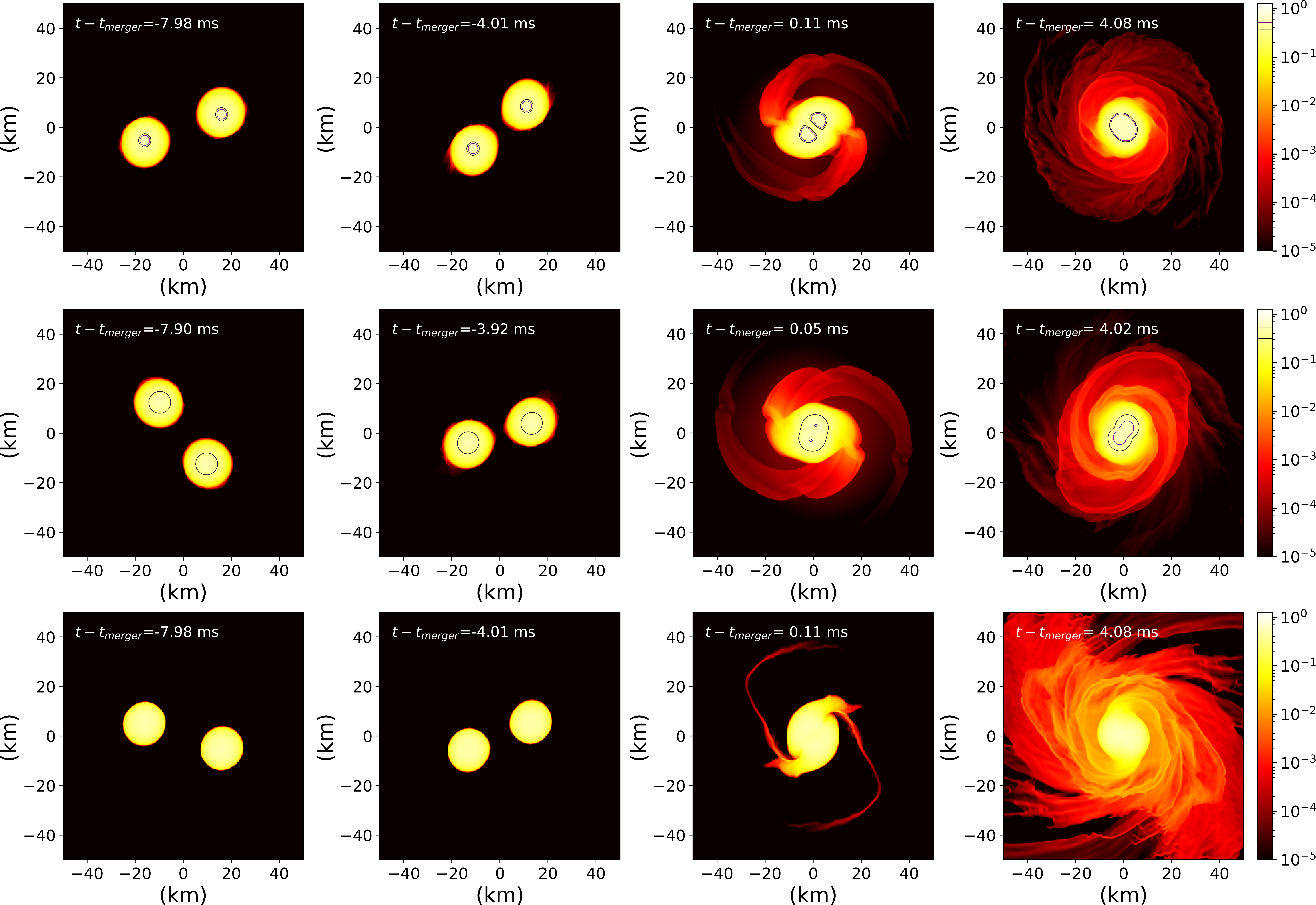}
	\caption{Density distributions with timestamps $\Delta t = t-t_{\rm merger}\simeq -8, -4, 0, 4$~ms on the $x-y$ plane of the BNS simulations with MC (upper), GC (middle), and CC (lower) type transitions corresponding to GWs in Figure~\ref{fig:35-150}. The color scale is in units of fm$^{-3}$ with $n_0\simeq 0.16$~fm$^{-3}$. Contours in figures and color bars denote the exact densities where phase transitions start and end. }
	\label{fig:density}
\end{figure}
\end{center}

Figure~\ref{fig:density} illustrates the spatial density evolution resulting from BNS mergers employing different constructions of the transition. The top, middle, and bottom rows correspond to the simulations with MC, GC, and CC type transitions, respectively. From the density contours in the top row we can see that the CSS phase persists within the compact star cores throughout the entire BNS merger process, including the inspiral phase. In contrast, in the GC scenario depicted in the middle row, the CSS phase emerges in the post-merger phase once the density reaches a critical value. In the bottom row, i.e. crossover scenario, no distinct contours are present, as it is impossible to distinguish the exact density where the transition begins and ends.

\section{Conclusion and discussion}

We comprehensively analyzed the impacts of various types of the hadron-quark transition on the properties of the NSs and GWs emitted during the BNS mergers. Explicitly, we compared the differences of the GW signals in time domain and frequency domain arising from the choices of the parameters in the EoS, the scenarios of the hadron-quark transition, and the masses of the component in the binary system. The main results can be summarized as follows:
\begin{enumerate}
	\item In MC, the GW signals are sensitive to the location of the phase transition and the latent heat. The smaller the transition density, the larger the GW frequency in the post-merger phase, the shorter the inspiral period and the smaller frequency of the frequency amplitude peak. The lager the latent heat, the bigger the GW frequency in the post-merger phase and the shorter the post-merger period. The latent heat does not affect the frequency amplitude significantly. The same conclusions are found in the GC type transition. For the CC type transition, the inspiral period is sensitive to the hadron-quark transition parameter $p_{\rm tr}$ in the CSS parameterization of quark matter. 
	\item The GW signals are sensitive to the size of the quark core of NS. The bigger the quark core, the shorter the inspiral period, the larger the GW frequency in the post-merger phase. 
	\item A comparison of the GW signals emitted by using the three types of constructions shows that in the post-merger phase the frequency of GW increases with the evolution in the MC-type transition but is not sensitive in the other two types of transitions. The radiated energy is the biggest in GC-type transition and the smallest in CC-type transition.
\end{enumerate}

The findings indicate that the features of the time-frequency, and the duration of the inspiral, the peak frequency and the post-merger phases serve as distinguishing factors to identify the type of transition, in addition to the precise GWs. By integrating present results of GW signals, electromagnetic signals from other observations, and the future GW detection, there is potential to differentiate between various types of hadron-quark transitions.

\begin{center}
	\begin{figure}[htbp]
		\includegraphics[width=1\linewidth]{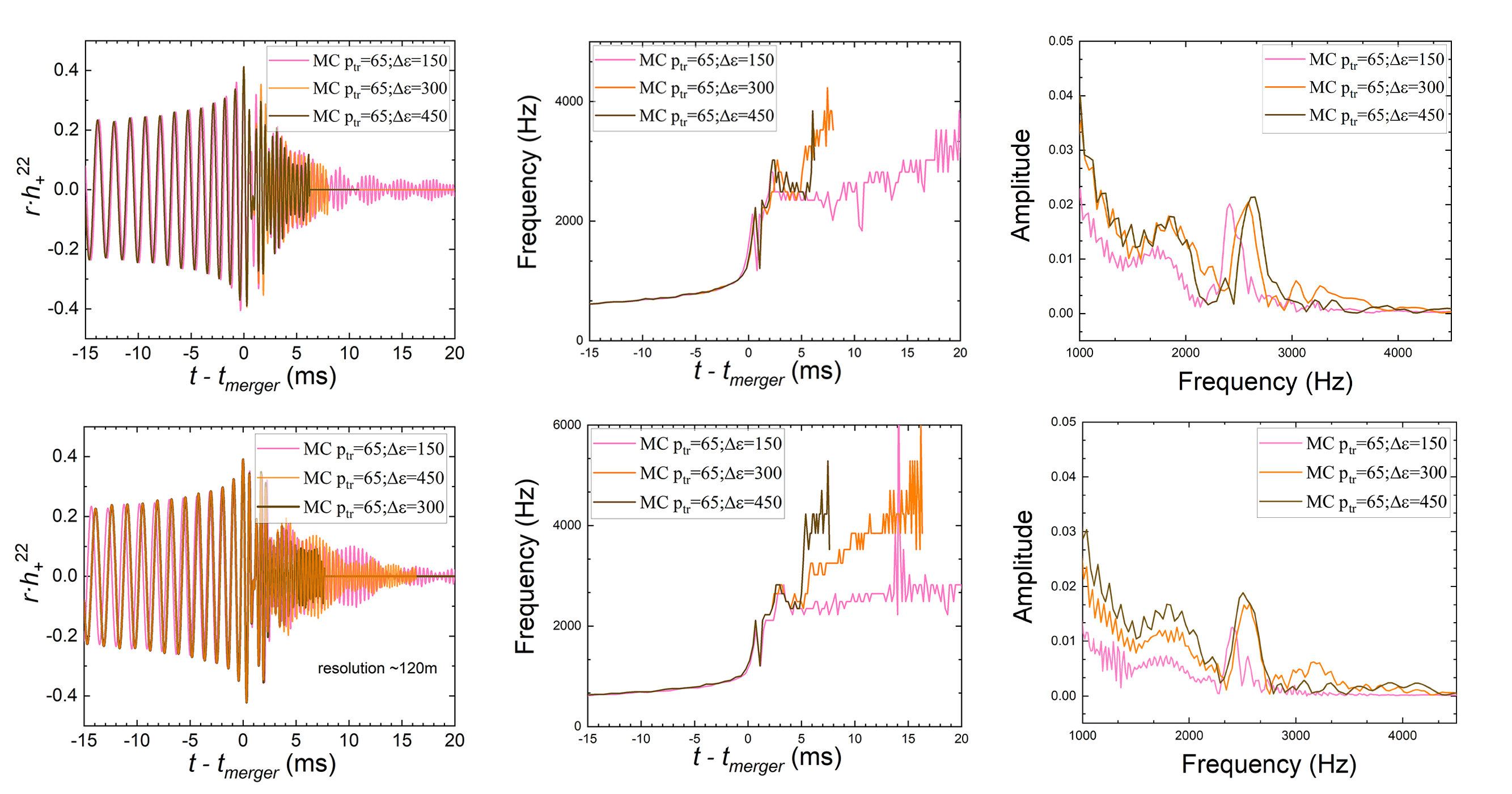}
		\caption{Gravitational waves from the BNS mergers with MC-type transition with resolution $260$~m (upper panel) and $120$m (lower panel). $p_{\rm tr}$ and $\Delta\varepsilon$ are in units of MeV/fm$^{3}$.}
		\label{fig:mcresolution}
	\end{figure}
	\end{center}

In the above numerical simulation, we took the typical resolution $260$m and used the fixed mesh grid. To check the convergence of the simulation and the uncertainties arising from the resolution, we increase the resolution to $120$m. The results illustrated in Figure~\ref{fig:mcresolution} show that although there are tiny differences in the GW signals, the conclusions discussed above remain intact. This simple check indicates that the results of this work are not affected by the the numerical resolution. 

We only presented the $(2,2)$ mode which is the dominant and quasi-symmetric quadrupole radiation mode of the multipole expansion of the gravitation radiation since we focused on the equal-mass binary systems in the above discussion. In some merger processes such as the un-equal mass binary systems, some higher modes like the $(2,1)$ one may be also interesting since the symmetry of the equal mass binary system loses~\cite{Centrella:2010mx}. We leave the detailed discussion of these aspects to the future publication.

Although it is difficult to measure the GWs after merger in the BNS system at this moment, it is promising to set up the Neutron Star Extreme Matter Observatory (NEMO) to measure the kilohertz GW in the upcoming decade~\cite{Ackley:2020atn,Borhanian:2022czq}, as well as the potential detection in O5 or 3G~\cite{Torres-Rivas:2018svp}. Therefore, it is essential to conduct a through analysis of the available EoSs and set up a database for the future matching between the observational date and theoretical simulation. This process will enable the refinement and constraint of compact star models based on the matched results.

In this work, we only considered the zero-temperature EOS with the finite temperature effect implemented through the ideal, nonrelativistic fermion gas approximation~\cite{Keil:1995hw}. Although this approximation works well for hadronic matter, the deviation is significant for quark matter. The temperature dependence of the EOS and the phase transition boundaries is highly relevant for the BNS mergers and the GW signals stemming from the post-merger phase which might be measurable in the next generation observatories~\cite{Blacker:2023afl,Fields:2023bhs}. This dependence should be properly involved in the explicit model if one wants to pin down some physics by matching the model calculation with the observation data. Since we varied the type of the hadron-quark transitions and the model parameters for the purpose to investigate their effects on GW signals, the impact of the thermal effect is beyond the scope of this work and our conclusions above are intact.

We finally want to say that, as one can see from the present explicit calculation and existing literature, the lower the transition pressure/density, the softer the EoS and thereby the easier the compact quark matter object collapses to a black hole. In the early universe before hadronization, there maybe compact objects composed solely of quarks. As temperature decreases, these compact objects may either collapse to or annihilate to black holes therefore provide the origin of certain black holes~\cite{Harko:2009ysn,Heald:2018htu}. We will investigate this aspect in the future work.

\begin{acknowledgements}


The work of Y.L.M. was supported in part by the National Key R\&D Program of China under grant No. 2021YFC2202900 and the National Natural Science Foundation of China (NSFC) under grant Nos. 12347103, 11875147 and 12147103. The work of Y.L.W. was supported in part by the National Key Research and Development Program of China under grant No.2020YFC2201501, the National Natural Science Foundation of China (NSFC) under grants Nos. 12347103, 12147103, 11821505, and the Strategic Priority Research Program of the Chinese Academy of Sciences under grant No. XDB23030100.
\end{acknowledgements}

\bibliographystyle{raa}
\bibliography{RefPhaseTranGW}

\begin{thebibliography}{107}
\providecommand\natexlab[1]{#1}
\providecommand\JournalTitle[1]{#1}

\bibitem[Abbott {et~al.}(2017)]{LIGOScientific:2017vwq}
Abbott, B.~P., {et~al.} 2017, Phys. Rev. Lett., 119, 161101

\bibitem[Abrahams \& Evans(1990)]{Abrahams:1990jf}
Abrahams, A.~M., \& Evans, C.~R. 1990, Phys. Rev. D, 42, 2585

\bibitem[Abrahams \& Price(1996)]{Abrahams:1995gn}
Abrahams, A.~M., \& Price, R.~H. 1996, Phys. Rev. D, 53, 1963

\bibitem[Ackley {et~al.}(2020)]{Ackley:2020atn}
Ackley, K., {et~al.} 2020, Publ. Astron. Soc. Austral., 37, e047

\bibitem[Akmal {et~al.}(1998)]{Akmal:1998cf}
Akmal, A., Pandharipande, V.~R., \& Ravenhall, D.~G. 1998, Phys. Rev. C, 58,
  1804

\bibitem[Albright {et~al.}(2014)]{Albright:2014gva}
Albright, M., Kapusta, J., \& Young, C. 2014, Phys. Rev. C, 90, 024915

\bibitem[Albright {et~al.}(2015)]{Albright:2015uua}
Albright, M., Kapusta, J., \& Young, C. 2015, Phys. Rev. C, 92, 044904

\bibitem[Alford {et~al.}(2015)]{Alford:2015dpa}
Alford, M.~G., Burgio, G.~F., Han, S., Taranto, G., \& Zappal\`a, D. 2015,
  Phys. Rev. D, 92, 083002

\bibitem[Alford {et~al.}(2013)]{Alford:2013aca}
Alford, M.~G., Han, S., \& Prakash, M. 2013, Phys. Rev. D, 88, 083013

\bibitem[Alford \& Sedrakian(2017)]{Alford:2017qgh}
Alford, M.~G., \& Sedrakian, A. 2017, Phys. Rev. Lett., 119, 161104

\bibitem[Allen(1977)]{allen1977short}
Allen, J. 1977, IEEE transactions on acoustics, speech, and signal processing,
  25, 235

\bibitem[Allen \& Rabiner(1977)]{allen:1977unified}
Allen, J.~B., \& Rabiner, L.~R. 1977, Proceedings of the IEEE, 65, 1558

\bibitem[Alsing {et~al.}(2018)]{Alsing:2017bbc}
Alsing, J., Silva, H.~O., \& Berti, E. 2018, Mon. Not. Roy. Astron. Soc., 478,
  1377

\bibitem[Alvarez-Castillo \& Blaschke(2017)]{Alvarez-Castillo:2017qki}
Alvarez-Castillo, D.~E., \& Blaschke, D.~B. 2017, Phys. Rev. C, 96, 045809

\bibitem[Annala {et~al.}(2020)]{Annala:2019puf}
Annala, E., Gorda, T., Kurkela, A., N\"attil\"a, J., \& Vuorinen, A. 2020,
  Nature Phys., 16, 907

\bibitem[Annala {et~al.}(2018)]{Annala:2017llu}
Annala, E., Gorda, T., Kurkela, A., \& Vuorinen, A. 2018, Phys. Rev. Lett.,
  120, 172703

\bibitem[Anninos {et~al.}(1994)]{Anninos:1994vw}
Anninos, P., Hobill, D., Seidel, E., Smarr, L., \& Suen, W.-M. 1994,
  arXiv:gr-qc/9408042

\bibitem[Antoniadis {et~al.}(2013)]{Antoniadis:2013pzd}
Antoniadis, J., {et~al.} 2013, Science, 340, 6131

\bibitem[Arzoumanian {et~al.}(2018)]{NANOGrav:2017wvv}
Arzoumanian, Z., {et~al.} 2018, Astrophys. J. Suppl., 235, 37

\bibitem[Baiotti {et~al.}(2005)]{Baiotti:2004wn}
Baiotti, L., Hawke, I., Montero, P.~J., {et~al.} 2005, Phys. Rev. D, 71, 024035

\bibitem[Bauswein {et~al.}(2019)]{Bauswein:2018bma}
Bauswein, A., Bastian, N.-U.~F., Blaschke, D.~B., {et~al.} 2019, Phys. Rev.
  Lett., 122, 061102

\bibitem[Bauswein \& Blacker(2020)]{Bauswein:2020ggy}
Bauswein, A., \& Blacker, S. 2020, Eur. Phys. J. ST, 229, 3595

\bibitem[Bauswein {et~al.}(2010)]{Bauswein:2010dn}
Bauswein, A., Janka, H.~T., \& Oechslin, R. 2010, Phys. Rev. D, 82, 084043

\bibitem[Baym {et~al.}(2019)]{Baym:2019iky}
Baym, G., Furusawa, S., Hatsuda, T., Kojo, T., \& Togashi, H. 2019, Astrophys.
  J., 885, 42

\bibitem[Baym {et~al.}(2018)]{Baym:2017whm}
Baym, G., Hatsuda, T., Kojo, T., {et~al.} 2018, Rept. Prog. Phys., 81, 056902

\bibitem[Benic {et~al.}(2015)]{Benic:2014jia}
Benic, S., Blaschke, D., Alvarez-Castillo, D.~E., Fischer, T., \& Typel, S.
  2015, Astron. Astrophys., 577, A40

\bibitem[Bhattacharyya {et~al.}(2010)]{Bhattacharyya:2009fg}
Bhattacharyya, A., Mishustin, I.~N., \& Greiner, W. 2010, J. Phys. G, 37,
  025201

\bibitem[Blacker {et~al.}(2023)]{Blacker:2023afl}
Blacker, S., Bauswein, A., \& Typel, S. 2023, Phys. Rev. D, 108, 063032

\bibitem[Borhanian \& Sathyaprakash(2022)]{Borhanian:2022czq}
Borhanian, S., \& Sathyaprakash, B.~S. 2022, arXiv:2202.11048

\bibitem[Borsanyi {et~al.}(2010)]{Borsanyi:2010cj}
Borsanyi, S., Endrodi, G., Fodor, Z., {et~al.} 2010, JHEP, 11, 077

\bibitem[Brown \& Rho(2002)]{Brown:2001nh}
Brown, G.~E., \& Rho, M. 2002, Phys. Rept., 363, 85

\bibitem[Centrella {et~al.}(2010)]{Centrella:2010mx}
Centrella, J., Baker, J.~G., Kelly, B.~J., \& van Meter, J.~R. 2010, Rev. Mod.
  Phys., 82, 3069

\bibitem[Chatziioannou \& Han(2020)]{Chatziioannou:2019yko}
Chatziioannou, K., \& Han, S. 2020, Phys. Rev. D, 101, 044019

\bibitem[Chen(2015)]{Chen:2015gba}
Chen, L.-W. 2015, EPJ Web Conf., 88, 00017

\bibitem[Cromartie {et~al.}(2019)]{NANOGrav:2019jur}
Cromartie, H.~T., {et~al.} 2019, Nature Astron., 4, 72

\bibitem[Damour \& Nagar(2010)]{Damour:2009wj}
Damour, T., \& Nagar, A. 2010, Phys. Rev. D, 81, 084016

\bibitem[De~Pietri {et~al.}(2016)]{DePietri:2015lya}
De~Pietri, R., Feo, A., Maione, F., \& L\"offler, F. 2016, Phys. Rev. D, 93,
  064047

\bibitem[Demorest {et~al.}(2010)]{Demorest:2010bx}
Demorest, P., Pennucci, T., Ransom, S., Roberts, M., \& Hessels, J. 2010,
  Nature, 467, 1081

\bibitem[Drews \& Weise(2017)]{Drews:2016wpi}
Drews, M., \& Weise, W. 2017, Prog. Part. Nucl. Phys., 93, 69

\bibitem[Endo {et~al.}(2005)]{Endo:2005va}
Endo, T., Maruyama, T., Chiba, S., \& Tatsumi, T. 2005, in {Hot Points in
  Astrophysics and Cosmology}: {Helmholtz International Summer School and
  Workshop}, 89

\bibitem[Ferreira {et~al.}(2021)]{Ferreira:2021osk}
Ferreira, M., C\^amara~Pereira, R., \& Provid\^encia, C. 2021, Phys. Rev. D,
  103, 123020

\bibitem[Fields {et~al.}(2023)]{Fields:2023bhs}
Fields, J., Prakash, A., Breschi, M., {et~al.} 2023, Astrophys. J. Lett., 952,
  L36

\bibitem[Fischer {et~al.}(2018)]{Fischer:2017lag}
Fischer, T., Bastian, N.-U.~F., Wu, M.-R., {et~al.} 2018, Nature Astron., 2,
  980

\bibitem[Fonseca {et~al.}(2016)]{Fonseca:2016tux}
Fonseca, E., {et~al.} 2016, Astrophys. J., 832, 167

\bibitem[Fonseca {et~al.}(2021)]{Fonseca:2021wxt}
Fonseca, E., {et~al.} 2021, Astrophys. J. Lett., 915, L12

\bibitem[Freedman \& McLerran(1977)]{Freedman:1976ub}
Freedman, B.~A., \& McLerran, L.~D. 1977, Phys. Rev. D, 16, 1169

\bibitem[Fujimoto {et~al.}(2023)]{Fujimoto:2022xhv}
Fujimoto, Y., Fukushima, K., Hotokezaka, K., \& Kyutoku, K. 2023, Phys. Rev.
  Lett., 130, 091404

\bibitem[Fukushima \& Hatsuda(2011)]{Fukushima:2010bq}
Fukushima, K., \& Hatsuda, T. 2011, Rept. Prog. Phys., 74, 014001

\bibitem[Fukushima {et~al.}(2020)]{Fukushima:2020cmk}
Fukushima, K., Kojo, T., \& Weise, W. 2020, Phys. Rev. D, 102, 096017

\bibitem[Fukushima \& Sasaki(2013)]{Fukushima:2013rx}
Fukushima, K., \& Sasaki, C. 2013, Prog. Part. Nucl. Phys., 72, 99

\bibitem[Gil {et~al.}(2019)]{Gil:2018yah}
Gil, H., Papakonstantinou, P., Hyun, C.~H., \& Oh, Y. 2019, Phys. Rev. C, 99,
  064319

\bibitem[Glendenning \& Kettner(2000)]{Glendenning:1998ag}
Glendenning, N.~K., \& Kettner, C. 2000, Astron. Astrophys., 353, L9

\bibitem[Glendenning \& Moszkowski(1991)]{Glendenning:1991es}
Glendenning, N.~K., \& Moszkowski, S.~A. 1991, Phys. Rev. Lett., 67, 2414

\bibitem[Gourgoulhon {et~al.}(2001)]{Gourgoulhon:2000nn}
Gourgoulhon, E., Grandclement, P., Taniguchi, K., Marck, J.-A., \& Bonazzola,
  S. 2001, Phys. Rev. D, 63, 064029

\bibitem[Han \& Prakash(2020)]{Han:2020adu}
Han, S., \& Prakash, M. 2020, Astrophys. J., 899, 164

\bibitem[Han \& Steiner(2019)]{Han:2018mtj}
Han, S., \& Steiner, A.~W. 2019, Phys. Rev. D, 99, 083014

\bibitem[Harko {et~al.}(2009)]{Harko:2009ysn}
Harko, T., Cheng, K.~S., \& Kovacs, Z. 2009, Mon. Not. Roy. Astron. Soc., 400,
  1632

\bibitem[Hawke {et~al.}(2005)]{Hawke:2005zw}
Hawke, I., Loffler, F., \& Nerozzi, A. 2005, Phys. Rev. D, 71, 104006

\bibitem[Heald(2018)]{Heald:2018htu}
Heald, G. 2018, Phys. Essays, 31, 449

\bibitem[Holt {et~al.}(2016)]{Holt:2014hma}
Holt, J.~W., Rho, M., \& Weise, W. 2016, Phys. Rept., 621, 2

\bibitem[Hotokezaka {et~al.}(2013)]{Hotokezaka:2013iia}
Hotokezaka, K., Kiuchi, K., Kyutoku, K., {et~al.} 2013, Phys. Rev. D, 88,
  044026

\bibitem[Hotokezaka {et~al.}(2011)]{Hotokezaka:2011dh}
Hotokezaka, K., Kyutoku, K., Okawa, H., Shibata, M., \& Kiuchi, K. 2011, Phys.
  Rev. D, 83, 124008

\bibitem[Janka {et~al.}(1993)]{1993AA268360J}
Janka, H.~T., Zwerger, T., \& Moenchmeyer, R. 1993, Astron. Astrophys., 268,
  360

\bibitem[Kalogera \& Baym(1996)]{Kalogera:1996ci}
Kalogera, V., \& Baym, G. 1996, Astrophys. J. Lett., 470, L61

\bibitem[Kanakis-Pegios {et~al.}(2020)]{Kanakis-Pegios:2020jnf}
Kanakis-Pegios, A., Koliogiannis, P.~S., \& Moustakidis, C.~C. 2020, Phys. Rev.
  C, 102, 055801

\bibitem[Kanakis-Pegios {et~al.}(2021)]{Kanakis-Pegios:2020kzp}
Kanakis-Pegios, A., Koliogiannis, P.~S., \& Moustakidis, C.~C. 2021, Symmetry,
  13, 183

\bibitem[Kapusta \& Welle(2021)]{Kapusta:2021ney}
Kapusta, J.~I., \& Welle, T. 2021, Phys. Rev. C, 104, L012801

\bibitem[Keil \& Janka(1995)]{Keil:1995hw}
Keil, W., \& Janka, H.~T. 1995, Astron. Astrophys., 296, 145

\bibitem[Klahn {et~al.}(2006)]{Klahn:2006ir}
Klahn, T., {et~al.} 2006, Phys. Rev. C, 74, 035802

\bibitem[Kojo {et~al.}(2022)]{Kojo:2021wax}
Kojo, T., Baym, G., \& Hatsuda, T. 2022, Astrophys. J., 934, 46

\bibitem[Kojo {et~al.}(2015)]{Kojo:2014rca}
Kojo, T., Powell, P.~D., Song, Y., \& Baym, G. 2015, Phys. Rev. D, 91, 045003

\bibitem[Koranda {et~al.}(1997)]{Koranda:1996jm}
Koranda, S., Stergioulas, N., \& Friedman, J.~L. 1997, Astrophys. J., 488, 799

\bibitem[Lalazissis {et~al.}(1997)]{Lalazissis:1996rd}
Lalazissis, G.~A., Konig, J., \& Ring, P. 1997, Phys. Rev. C, 55, 540

\bibitem[Li {et~al.}(2019)]{Li:2019xxz}
Li, B.-A., Krastev, P.~G., Wen, D.-H., \& Zhang, N.-B. 2019, Eur. Phys. J. A,
  55, 117

\bibitem[Lim {et~al.}(2017)]{Lim:2015lia}
Lim, Y., Hyun, C.~H., \& Lee, C.-H. 2017, Int. J. Mod. Phys. E, 26, 1750015

\bibitem[Loffler {et~al.}(2012)]{Loffler:2011ay}
Loffler, F., {et~al.} 2012, Class. Quant. Grav., 29, 115001

\bibitem[Lopes(2022)]{Lopes:2021zfe}
Lopes, L.~L. 2022, Commun. Theor. Phys., 74, 015302

\bibitem[Lovato {et~al.}(2022)]{Lovato:2022vgq}
Lovato, A., {et~al.} 2022, arXiv:2211.02224

\bibitem[Ma \& Rho(2020)]{Ma:2019ery}
Ma, Y.-L., \& Rho, M. 2020, Prog. Part. Nucl. Phys., 113, 103791

\bibitem[Maione {et~al.}(2016)]{Maione:2016zqz}
Maione, F., De~Pietri, R., Feo, A., \& L\"offler, F. 2016, Class. Quant. Grav.,
  33, 175009

\bibitem[Margaritis {et~al.}(2020)]{Margaritis:2019hfq}
Margaritis, C., Koliogiannis, P.~S., \& Moustakidis, C.~C. 2020, Phys. Rev. D,
  101, 043023

\bibitem[Masuda {et~al.}(2013{\natexlab{a}})]{Masuda:2012kf}
Masuda, K., Hatsuda, T., \& Takatsuka, T. 2013{\natexlab{a}}, Astrophys. J.,
  764, 12

\bibitem[Masuda {et~al.}(2013{\natexlab{b}})]{Masuda:2012ed}
Masuda, K., Hatsuda, T., \& Takatsuka, T. 2013{\natexlab{b}}, PTEP, 2013,
  073D01

\bibitem[McLerran \& Reddy(2019)]{McLerran:2018hbz}
McLerran, L., \& Reddy, S. 2019, Phys. Rev. Lett., 122, 122701

\bibitem[Mondal {et~al.}(2023)]{Mondal:2023gbf}
Mondal, C., Antonelli, M., Gulminelli, F., {et~al.} 2023, Mon. Not. Roy.
  Astron. Soc., 524, 3464

\bibitem[Montana {et~al.}(2019)]{Montana:2018bkb}
Montana, G., Tolos, L., Hanauske, M., \& Rezzolla, L. 2019, Phys. Rev. D, 99,
  103009

\bibitem[Most {et~al.}(2020)]{Most:2019onn}
Most, E.~R., Jens~Papenfort, L., Dexheimer, V., {et~al.} 2020, Eur. Phys. J. A,
  56, 59

\bibitem[Most {et~al.}(2019)]{Most:2018eaw}
Most, E.~R., Papenfort, L.~J., Dexheimer, V., {et~al.} 2019, Phys. Rev. Lett.,
  122, 061101

\bibitem[M\"osta {et~al.}(2014)]{Mosta:2013gwu}
M\"osta, P., Mundim, B.~C., Faber, J.~A., {et~al.} 2014, Class. Quant. Grav.,
  31, 015005

\bibitem[Orsaria {et~al.}(2019)]{Orsaria:2019ftf}
Orsaria, M.~G., Malfatti, G., Mariani, M., {et~al.} 2019, J. Phys. G, 46,
  073002

\bibitem[Oter {et~al.}(2019)]{Oter:2019rqp}
Oter, E.~L., Windisch, A., Llanes-Estrada, F.~J., \& Alford, M. 2019, J. Phys.
  G, 46, 084001

\bibitem[Prakash {et~al.}(2021)]{Prakash:2021wpz}
Prakash, A., Radice, D., Logoteta, D., {et~al.} 2021, Phys. Rev. D, 104, 083029

\bibitem[Radice {et~al.}(2017)]{Radice:2016rys}
Radice, D., Bernuzzi, S., Del~Pozzo, W., Roberts, L.~F., \& Ott, C.~D. 2017,
  Astrophys. J. Lett., 842, L10

\bibitem[Read {et~al.}(2009)]{Read:2008iy}
Read, J.~S., Lackey, B.~D., Owen, B.~J., \& Friedman, J.~L. 2009, Phys. Rev. D,
  79, 124032

\bibitem[Riley {et~al.}(2019)]{Riley:2019yda}
Riley, T.~E., {et~al.} 2019, Astrophys. J. Lett., 887, L21

\bibitem[Riley {et~al.}(2021)]{Riley:2021pdl}
Riley, T.~E., {et~al.} 2021, Astrophys. J. Lett., 918, L27

\bibitem[Sarin {et~al.}(2020)]{Sarin:2020pwr}
Sarin, N., Lasky, P.~D., \& Ashton, G. 2020, Phys. Rev. D, 101, 063021

\bibitem[Schnetter {et~al.}(2004)]{Schnetter:2003rb}
Schnetter, E., Hawley, S.~H., \& Hawke, I. 2004, Class. Quant. Grav., 21, 1465

\bibitem[Sejdi{\'c} {et~al.}(2009)]{sejdic2009time}
Sejdi{\'c}, E., Djurovi{\'c}, I., \& Jiang, J. 2009, Digital signal processing,
  19, 153

\bibitem[Sekiguchi {et~al.}(2011)]{Sekiguchi:2011mc}
Sekiguchi, Y., Kiuchi, K., Kyutoku, K., \& Shibata, M. 2011, Phys. Rev. Lett.,
  107, 211101

\bibitem[Subedi {et~al.}(2008)]{Subedi:2008zz}
Subedi, R., {et~al.} 2008, Science, 320, 1476

\bibitem[Tews {et~al.}(2018)]{Tews:2018iwm}
Tews, I., Margueron, J., \& Reddy, S. 2018, Phys. Rev. C, 98, 045804

\bibitem[Torres-Rivas {et~al.}(2019)]{Torres-Rivas:2018svp}
Torres-Rivas, A., Chatziioannou, K., Bauswein, A., \& Clark, J.~A. 2019, Phys.
  Rev. D, 99, 044014

\bibitem[Tsang {et~al.}(2018)]{Tsang:2018kqj}
Tsang, C.~Y., Tsang, M.~B., Danielewicz, P., Lynch, W.~G., \& Fattoyev, F.~J.
  2018, arXiv:1807.06571

\bibitem[Yang {et~al.}(2021)]{Yang:2020ucv}
Yang, W.-C., Ma, Y.-L., \& Wu, Y.-L. 2021, Sci. China Phys. Mech. Astron., 64,
  252011

\bibitem[Zdunik \& Haensel(2013)]{Zdunik:2012dj}
Zdunik, J.~L., \& Haensel, P. 2013, Astron. Astrophys., 551, A61

\bibitem[Zilh\~ao \& L\"offler(2013)]{Zilhao:2013hia}
Zilh\~ao, M., \& L\"offler, F. 2013, Int. J. Mod. Phys. A, 28, 1340014

\end{thebibliography}

\end{document}